RESEARCH ARTICLE

# A convolutional neural network deep learning method for model class selection

**Marios Impraimakis**

Department of Civil, Maritime and Environmental Engineering, University of Southampton, Southampton, UK

**Correspondence**
Marios Impraimakis, Department of Civil, Maritime and Environmental Engineering, University of Southampton, Southampton SO16 7QF, UK.
Email: m.impraimakis@southampton.ac.uk

**Abstract**

The response-only model class selection capability of a novel deep convolutional neural network method is examined herein in a simple, yet effective, manner. Specifically, the responses from a unique degree of freedom along with their class information train and validate a one-dimensional convolutional neural network. In doing so, the network selects the model class of new and unlabeled signals without the need of the system input information, or full system identification. An optional physics-based algorithm enhancement is also examined using the Kalman filter to fuse the system response signals using the kinematics constraints of the acceleration and displacement data. Importantly, the method is shown to select the model class in slight signal variations attributed to the damping behavior or hysteresis behavior on both linear and nonlinear dynamic systems, as well as on a 3D building finite element model, providing a powerful tool for structural health monitoring applications.

**KEYWORDS**

artificial neural networks, convolutional neural networks, machine learning, model class selection-assessment, pattern recognition, physics-enhanced deep learning, structural health monitoring

## 1 | INTRODUCTION

The model class selection is an integral part of the system identification and monitoring process given that either the analytical or the numerical system models are, inevitably, only an approximation of the real system. It is particularly useful for engineering systems since it is difficult to be determined solely by the physics due to their empirical nature.

The importance of the model class selection, specifically, is highlighted by the fact that a more complicated model fits the data better than one which has fewer adjustable uncertain parameters, but it is likely results in data over-fitting and poor future predictions. This is attributed to the parameter fitting which depend too much on the detail of the data and the measurement noise.

To address those challenges, a long history of approaches exists. Akaike[1] introduced a likelihood function which penalizes the parameterization of the models, and Grigoriu et al.[2] suggested to penalize the complicated models over the simpler ones. Beck and Yuen[3] proposed the ranking of the model classes based on their response conditional probabilities which are calculated by the Bayes' theorem and the asymptotic expansion of each model class evidence, while Katafygiotis and Beck[4] introduced an algorithm to investigate the model identifiability in structural model updating using a network of trajectories which finds all other output-equivalent optimal models. Importantly, Ching and Chen[5] developed a simulation-based approach for the simultaneous Bayesian model updating, model class selection, and model averaging. Muto and Beck[6] implemented, later, the transitional Markov Chain Monte Carlo method for nonlinear structures under seismic loading. Additionally, Cheung and Beck[7] proposed a general method for calculating the model evidence based on the posterior samples of the Markov Chain Monte Carlo approach, while Beck[8] investigated the Laplace's method of asymptotic approximation and the Markov Chain Monte Carlo methods for a structural health monitoring benchmark problem.

Furthermore, Raftery et al.[9] developed the method of dynamic model averaging for online model class selection. Chatzi et al.[10] proposed and experimentally validated a twofold criterion based on the smoothness of the parameter prediction and the accuracy of the estimation. Yuen and Mu[11] developed a novel model class selection component into the extended Kalman filter algorithm, to simultaneously provide the model class selection and the parametric identification in a real-time manner. Importantly, Kontoroupi and Smyth[12] explored how the Bayesian model selection and the unscented Kalman filter scheme for joint state and parameter estimation can be integrated into a single method using each model's probability-plausibility computation. The Bayesian model class selection and the unscented Kalman filter joint scheme with the penalty-type Kullback–Leibler divergence was also investigated,[13] and the research is still ongoing.[14–25]

However, the current model class selection methodologies, apart from the class selection, incorporate also the system identification for each model. The main challenge here is derived from the effort of performing this task for partial unobservable systems, such as large systems under very limited information, or systems with unknown inputs. Similarly, this task is not trivial in empirical systems with nonlinear behavior where no acceptable closed-form equation representation exists.

A way to address those challenges is examined here using a generalized response-only and (after the training) real-time procedure based on the deep learning capabilities which selects automatically the system model class without having to identify its parameters, measure and estimate all dynamic states, or knowing the system input. The convolutional neural network approach is therefore employed of the deep learning library of methods. Importantly, the convolutional neural networks have already shown an impressive performance on selecting the class of visual imagery data[26] via an ability to recognize patterns. Here, a one-dimensional version is examined for the vibration signals, which has shown a great potentially for damage detection in one or more dimensions.[27–44] The ability to provide the model class selection using a unique degree of freedom (DOF) response measurement, without system identification, and by using a neural network classification approach makes this approach distinctive from the current methodologies.

The methodology, specifically, results in a fast and accurate nonparametric vibration-based tool for model class selection which directly classify the model based solely on response signals. An algorithm enchantment is also investigated when the dynamic state estimates of a Kalman filter as developed by Smyth and Wu[45] and implemented as a physics-enhanced kinematics constraint,[46] train a network to recognize their patterns and classify the new and unlabeled signals. In this way, the advantages of the Kalman filtering[47–55] are explored to improve the performance of the convolutional neural network. Due to the convolutional neural network ability to learn and extract the optimal features with a proper training, the proposed approach achieves an impressive model class selection accuracy despite the response-only nature of the signals.

The work is organized as follows: the Bayesian model class selection and the limitations are overviewed in Section 2. In Section 3, the standard convolutional neural network architecture is provided, as well as a comparison of the one-dimensional and the multi-dimensional convolutional neural network versions with a focus on the model class selection. In Section 4, the Kalman filter fusion is formulated for response-only, unknown input, and unknown model class systems. Section 5 provides the summary and the detailed algorithmic tables. Importantly, Sections 6, 7, and 8 investigate numerical applications on both linear and nonlinear dynamic systems, as well as on a 3D building finite element model. Subsequently, Section 9 presents a discussion, future research suggestions, and sensitivity analysis for the training process. Finally, the conclusions are provided in Section 10.

## 2 | BAYESIAN MODEL CLASS SELECTION

To select the model class $\mathbb{M}_i$ in a Bayesian framework, one needs to use their prior probability distribution, and then assess their posterior probability plausibility. Let $\mathbb{M}$ be the space of the models $\mathbb{M}_{i:i_{max}}$. The posterior probability $P(\mathbb{M}_i \mid \mathbf{y}, \mathbb{M})$ of the model class $\mathbb{M}_i$ is defined using the Bayes theorem as:

$$P(\mathbb{M}_i \mid \mathbf{y}, \mathbb{M}) = \frac{p(\mathbf{y} \mid \mathbb{M}_i) \cdot P(\mathbb{M}_i \mid \mathbb{M})}{p(\mathbf{y} \mid \mathbb{M})} \tag{1}$$

where, $P(\mathbb{M}_i \mid \mathbb{M})$ is the prior probability of $\mathbb{M}_i$, $\mathbf{y}$ is the measurement vector, and $p(\mathbf{y} \mid \mathbb{M}_i)$ is the evidence given the model $\mathbb{M}_i$. The denominator is replaced by the summation of the prior probability and the likelihood for every model class, written as:

$$P(\mathbb{M}_i \mid \mathbf{y}, \mathbb{M}) = \frac{p(\mathbf{y} \mid \mathbb{M}_i) \cdot P(\mathbb{M}_i \mid \mathbb{M})}{\sum_i^{i_{max}} \left\{ p(\mathbf{y} \mid \mathbb{M}_i) \cdot P(\mathbb{M}_i \mid \mathbb{M}) \right\}} \tag{2}$$

Let $\theta_j \in \mathbb{M}_i$ be the parameter $j$ of the model $\mathbb{M}_i$. The posterior probability distribution $p(\theta_j \mid \mathbf{y}, \mathbb{M}_i)$ of $\theta_j$ is written as:

$$p(\theta_j \mid \mathbf{y}, \mathbb{M}_i) = \frac{p(\mathbf{y} \mid \theta_j, \mathbb{M}_i) \cdot p(\theta_j \mid \mathbb{M}_i)}{\int_\theta p(\mathbf{y} \mid \theta_j, \mathbb{M}_i) \cdot p(\theta_j \mid \mathbb{M}_i)\, d\theta} = \frac{p(\mathbf{y} \mid \theta_j, \mathbb{M}_i) \cdot p(\theta_j \mid \mathbb{M}_i)}{p(\mathbf{y} \mid \mathbb{M}_i)} \tag{3}$$

where, $p(\mathbf{y} \mid \theta_j, \mathbb{M}_i)$ is the likelihood given the parameter $\theta_j$ and the model $\mathbb{M}_i$, and $p(\theta_j \mid \mathbb{M}_i)$ is the prior probability density function of $\theta_j$ given the model $\mathbb{M}_i$. Here, computing the evidence $p(\mathbf{y} \mid \mathbb{M}_i)$ for each model $\mathbb{M}_i$ is not trivial. Specifically, the high-dimensional integral is usually analytically intractable, for instance when nonconjugate prior probabilities and/or latent variables exist.

To this end, stochastic simulation methods are used. Particularly, the Markov chain Monte Carlo methods generate samples from the posterior distribution, and then compute the likelihood using the following identity of a rearranged Bayes theorem for every $\theta_j$:

$$ln(p(\mathbf{y} \mid \mathbb{M}_i)) = ln\big(p(\mathbf{y} \mid \theta_j, \mathbb{M}_i)\big) + ln\big(p(\theta_j \mid \mathbb{M}_i)\big) - ln\big(p(\theta_j \mid \mathbf{y}, \mathbb{M}_i)\big) \tag{4}$$

where, the natural logarithm $ln(\bullet)$ is applied to avoid numerical overflows. Equation (4) is also written as[6]:

$$ln(p(\mathbf{y} \mid \mathbb{M}_i)) = \int_\theta ln\big(p(\mathbf{y} \mid \theta_j, \mathbb{M}_i)\big)\, p(\theta_j \mid \mathbf{y}, \mathbb{M}_i)\, \boldsymbol{d\theta} - \int_\theta ln\left(\frac{p(\theta_j \mid \mathbf{y}, \mathbb{M}_i)}{p(\theta_j \mid \mathbb{M}_i)}\right) p(\theta_j \mid \mathbf{y}, \mathbb{M}_i)\, \boldsymbol{d\theta} \tag{5}$$

where, the first expectation term measures the posterior average data fit of the parameter set $\mathbb{M}_i$, while the penalty-type second one represents the Kullback–Leibler divergence[56] between the parameter posterior and prior probability distributions.

Finally, the identification with the highest evidence $ln(p(\mathbf{y} \mid \mathbb{M}_i))$[12] or the least Kullback–Leibler divergence[13] is selected as the one with the most plausible model class.

However, this approach requires a parametric model-based implementation of the model class selection, which inevitably require a parameter estimation and the system input knowledge for input-output identification. Contrastingly in the convolutional neural network approach, a response-only nonparametric signal-based approach is implemented by using the machine learning means to directly select the model class by recognizing signal patterns.

## 3 | CONVOLUTIONAL NEURAL NETWORK ARCHITECTURE

The convolutional neural networks are a type of deep learning artificial neural network methods with an ability to recognize patterns in visual data. They are composed of multiple building blocks which automatically and adaptively learn spatial hierarchies of features.

The one-dimensional convolutional neural networks (1D CNN) have been proven to be highly effective in a variety of signal processing tasks. The fundamental building block of a 1D CNN is the convolutional layer. The convolutional layer applies a set of filters to the input signal, producing a set of feature maps. The filters have a fixed size and slide over the input signal, computing a dot product at each location. In doing so, the resulting feature maps capture different aspects of the input signal, such as local trends and patterns. In practice, a 1D CNN may have multiple convolutional layers with different filter sizes and number of filters. Each layer can apply a different set of filters to the input signal, allowing the network to capture different aspects of the signal at different scales.

The examined one-dimensional convolutional neural network compares to the multi-dimensional counterparts as follows. A one-dimensional configuration fuses the feature extraction and the learning phases of the dynamic states. One-dimensional arrays are used instead of two-dimensional matrices for both the kernels and the feature maps.

Additionally, the network architecture has the hidden neurons of the convolution layers which perform both the convolution and the sub-sampling operations. The fully-connected layers are identical to the hidden layers of the multi-layer perceptrons where the classification task is mainly realized. Accordingly, the multi-dimensional matrix manipulations, namely the convolution and the lateral rotation, are replaced by their one-dimensional counterparts, namely the convolution and the reverse operations. Finally, the parameters for the kernel size and the sub-sampling are scalars. Importantly, this simplified structure of the convolution neural network requires only one-dimensional convolutions and therefore, a mobile and low-cost hardware implementation for near real-time applications. The algorithmic details of the 1D CNN are provided in Section 5.

A short description of the additional layers in the convolutional neural network architecture is provided. **Input Layer**: The layer where the input is specified. **Convolutional Layer**: The layer where the filters are applied to the input, usually between a subarray of the input array and the filter, and where the neurons connect to the input subarray. In this layer, the number of feature maps is also determined. **Batch Normalization Layer**: The layer where the normalization of the activations and gradients occurs leading to a simpler optimization training problem. It is usually followed by a nonlinear activation function. **Pooling Layer**: The layer where the down-sampling operation is applied to reduce the spatial size of the feature map and to remove the redundant spatial information. This leads to an increase of the number of filters in deeper convolutional layers without increasing the required amount of computation per layer. **Fully Connected Layer**: The layer where the neurons connect to the neurons in the preceding layer to combine all the features learned by the previous layers and identify the larger patterns. Importantly, the last fully connected layer combines the features to classify the data and is equal to the number of classes in the input data. **Softmax Layer**: The layer where the activation function normalizes the output of the fully connected layer. The output of this layer consists of positive numbers that sum to one, which are then used as classification probabilities by the classification layer. **Classification Layer**: The final layer where the probabilities are returned by the activation function for each input to assign the mutually exclusive classes and compute the loss. Importantly, the training of the network is implemented usually by a stochastic gradient descent with a specified number of epochs, where an epoch is a full training cycle on the entire training data set.

In this work, the one-dimensional convolutional approach is applied to select the model class. The examined approach fuses both the feature extraction and the classification blocks into a single and compact learning body. The advantage is the ability to extract optimal model class-sensitive features automatically from the response-only signals.

## 4 | RESPONSE-ONLY AND UNKNOWN MODEL CLASS DYNAMIC STATE ESTIMATION USING THE KALMAN FILTER

For a further improvement of the 1D CNN performance with response-only signals when additional signals are available, the Kalman filter data fusion technique may be used by Smyth and Wu.[45,46] The Kalman filter algorithm, given a series of noisy measurements observed over time, estimates optimally the system dynamic states using a joint probability distribution over the states for each timeframe. The algorithm works in two steps: the first step is the prediction of the dynamic states using the dynamic process model which also propagates the uncertainty of the dynamic states. The second update step incorporates the measurements to calibrate the dynamic state estimation using a weighted average strategy, where more weight is given to the estimates with higher certainty. The algorithm is recursive and it is used online and, potentially, with real-time data.

Even for simple systems though, the knowledge of the system parameters and input is needed to predict future steps. This leads to an unavailability of filtering the signals when response-only and unknown model class scenarios are examined. To this end, the dynamic states are filtered using acceleration and displacement measurements[45,46] as:

$$\begin{Bmatrix} \dot{x} \\ \ddot{x} \end{Bmatrix} = \begin{bmatrix} 0 & 1 \\ 0 & 0 \end{bmatrix} \begin{Bmatrix} x \\ \dot{x} \end{Bmatrix} + \begin{Bmatrix} 0 \\ 1 \end{Bmatrix} a + \begin{Bmatrix} 0 \\ 1 \end{Bmatrix} \eta_a \tag{6}$$

$$y = d = \begin{bmatrix} 1 & 0 \end{bmatrix} \begin{Bmatrix} x \\ \dot{x} \end{Bmatrix} + \eta_d \tag{7}$$

where, $a$ and $d$ are the acceleration and displacement measurements, respectively, and $\eta_a$ and $\eta_d$ are their associated noise. It is assumed that $\eta_a$ and $\eta_d$ are white noise Gaussian processes. By introducing the state variables,

$$\mathbf{x} = \begin{Bmatrix} x_1 \\ x_2 \end{Bmatrix} = \begin{Bmatrix} x \\ \dot{x} \end{Bmatrix} \tag{8}$$

Equations (6) and (7), without the noise terms, are written in matrix form as:

$$\dot{\mathbf{x}} = \mathbf{A}\,\mathbf{x} + \mathbf{B}\,a \tag{9}$$

$$y = \mathbf{H}\,\mathbf{x} \tag{10}$$

If acceleration measurements are available at intervals of $\Delta t$, the process Equation (9) and the observation Equation (10) are discretized as:

$$\mathbf{x}(k+1) = \mathbf{A_d}\,\mathbf{x}(k) + \mathbf{B_d}\,a(k) \tag{11}$$

$$y(k+1) = \mathbf{H}\,\mathbf{x}(k+1) \tag{12}$$

namely,

$$\begin{Bmatrix} x_1(k+1) \\ x_2(k+1) \end{Bmatrix} = \begin{bmatrix} 1 & \Delta t \\ 0 & 1 \end{bmatrix} \begin{Bmatrix} x_1(k) \\ x_2(k) \end{Bmatrix} + \begin{Bmatrix} \Delta t^2/2 \\ \Delta t \end{Bmatrix} a(k) + \begin{Bmatrix} \Delta t^2/2 \\ \delta t \end{Bmatrix} \eta_a(k) \tag{13}$$

$$y = [1 \quad 0] \begin{bmatrix} x_1(k+1) \\ x_2(k+1) \end{bmatrix} + \eta_d(k+1) \tag{14}$$

where $k$ step stands for the $k \cdot \Delta t$ time instance.

In the approach of Equation (13), a physics-enhanced fusion of the displacement and the acceleration signals is implemented. Specifically, the object kinematics equation is employed as the system pseudo-model to provide the physical relationship between the displacement and acceleration data, without incorporating any knowledge of the actual system model and its class. Equation (13) is simpler written using the well-known body-motion equations as:

$$\text{Physical Kinematics Model } = \begin{cases} x_2 = x_2 + \dfrac{1}{2}\,a \cdot \Delta t \\ x_1 = x_1 + x_2 \cdot \Delta t + \dfrac{1}{2}\,a \cdot \Delta t^2 \end{cases} \tag{15}$$

where, the acceleration $a$ is assumed to be constant between each sequential steps; an assumption which does not lead to divergences due to the small value of $\Delta t$.

Overall, this fusion algorithm uses the Kalman filter which, given acceleration and displacement measurements, provides optimally the displacement and the velocity dynamic states. Importantly, the displacement measurement is provided by the integration of the acceleration signal on linear systems.

The obtained results can be used instead of raw signals to train, validate, and test the convolutional neural network for model class selection. Notably, those response-only signals are used as input data to the network, and they should not be confused with the output of the network.

Finally, the one-dimensional convolutional neural network procedure for model class selection is implemented as follows. The dynamic states of a unique system responses are loaded to train and validate the network. Importantly, these signals are already labeled with the model class. The one-dimensional convolutional neural network architecture is defined where the input size of the training data is specified as the number of their classes. Subsequently, the network training optimization algorithm is specified which included a mini-batch approach with an adequate number of epochs. For online purposes with a unique response training signal, the mini-batch size is set equal to 1, otherwise larger values also are used. Once the network is trained, it is used to evaluate the new and unlabeled signals, and select their model class. Importantly, no additional data such as the system input or the system parameters are needed.

## 5 | PROCEDURE SUMMARY

The overall procedure is illustrated here where each step is detailed in Table 1:

1. **Initialize the measurement filtering (optional for improved performance incorporating more data)**. Set the initial probability distributions for the dynamic states of each mode class response signal.

**TABLE 1** Kalman filter convolutional neural network (Kalman filter C-Net).

Step 1 (optional):

- Initialize the dynamic state estimation:
  - $k = 0$ (Time step)
  - $\mathbf{x_k} = \mathbb{E}[\mathbf{x_0}]$ ($\mathbb{E}$ stands for Expectation)
  - $\mathbf{P_k} = \mathbb{E}[(\mathbf{x_0} - \mathbf{x_k})(\mathbf{x_0} - \mathbf{x_k})^T]$ (Covariance matrix)

Step 2 (optional):

- Predict and estimate the dynamic states:
  - $\mathbf{x_{k+1}} = \mathbf{A_d}\,\mathbf{x_k} + \mathbf{B_d}\,a_k$ (prediction)
  - $\mathbf{P_{k+1}} = \mathbf{A_d}\,\mathbf{P_k}\,\mathbf{A_d^T} + \mathbf{Q_d}$
  - $\mathbf{J} = \mathbf{P_{k+1}}\,\mathbf{H^T}\,(\mathbf{R_d} + \mathbf{C}\,\mathbf{P_{k+1}}\,\mathbf{H^T})^{-1}$
  - $\mathbf{x_{k+1}} = \mathbf{x_{k+1}} + \mathbf{J}\,(\mathbf{y_{k+1}} - \mathbf{H}\,\mathbf{x_{k+1}})$ (estimation)
  - $\mathbf{P_{k+1}} = (\mathbf{I} - \mathbf{J}\,\mathbf{H})\,\mathbf{P_{k+1}}\,(\mathbf{I} - \mathbf{J}\,\mathbf{H})^T + \mathbf{J}\,\mathbf{R_d}\,\mathbf{J}^T$
- Repeat Step 2 for $k = k + 1$ until $k_{max}$

Step 3:

- Initialize randomly all weights for the neural network
- Forward propagate the input data:
  - $\mathbf{z_j^h} = b_j^h + \sum_{i=1}^{h-1} D(v_{ij}, \mathbf{s_i^{h-1}})$
  - $\mathbf{u_j^h} = F(\mathbf{z_j^h})$ and $\mathbf{s_j^h} = \mathbf{u_j^h}$ downsampling
- Compute the delta error at the output layer and back-propagate it:
  - $E = \sum_{h=1}^{N_h} (\mathbf{u_h^{N_h}} - \mathbf{r_h})^2$
  - $\partial E \,/\, \partial v_{ij}^{h-1} = \Delta_j^h\, \mathbf{u_i^{h-1}}$ and $\partial E \,/\, \partial b_j^h = \Delta_j^h$
  - $\partial E \,/\, \partial s_j^h = \sum_{h=1}^{N_h+1} \Delta_h^{h+1}\, v_{jh}$

Step 4:

- Post-process to compute the weight and bias sensitivities:
  - $\Delta_h^j = \partial E \,/\, \partial \mathbf{u_j^h} \cdot \partial \mathbf{u_j^h} \,/\, \partial \mathbf{z_j^h}$ (further back propagation)
- Update the weights and biases with the accumulation of sensitivities:
  - $v_{ij}^{h-1}(t+1) = v_{ij}^{h-1}(t) - \epsilon \cdot \partial E \,/\, \partial v_{ij}$ (adaptive $\epsilon \approx 0.001$)
  - $b_j^h(t+1) = b_j^h(t) - \epsilon \cdot \partial E \,/\, \partial b_j^h$

Step 5:

- Move to each next layer until the network is fully trained. Classify the unlabeled signals from Step 3 using the trained network.

2. **Filter the dynamic states online (optional for improved performance incorporating more data)**. Predict the dynamic states using the acceleration measurements and the discrete state-space modeling. Estimate the dynamic states using the displacement measurements. The displacement measurements may have a different rate than acceleration measurements.[45,46] Importantly, for linear systems double-integrate the acceleration measurements. Also, **Repeat the filtering for the full signal duration**. Repeat the Kalman filter procedure for all time steps to provide the full input.
3. **Feed the network**. Provide the one-dimensional convolutional neural network with the raw signals or filtered signals from Step 2 associated with their model class. At this point, generate randomly the weights of the network. Also, **Initialize the network training**. Start the network training where the signal data are propagated between the layers.
4. **Implement the back-propagation algorithm in the network training**. Post-process the signal data for the estimation of the weights and bias sensitivities. Update the weights and biases with the accumulation of sensitivities. **Finally, move to each next layer**.
5. **Select the model class**. Use the trained network to classify the unlabeled signals. Specifically, provide the new, unused, and unlabeled raw or filtered signals from Step 2 as an input to the network to output the model class.

**TABLE 2** Damping kernel functions $g(t)$ for the generalized damping model classes.

| Kernel function $g(t)$ | Constraints |
|---|---|
| • $\mu_1 e^{-\mu_1 t}$ | $\mu_1, t \geq 0$ |
| • $(\mu_2)^2\ t e^{-\mu_2 t}$ | $\mu_2, t \geq 0$ |
| • $2\sqrt{\frac{\mu_3}{\pi}} e^{-\mu_3 t^2}$ | $\mu_3, t \geq 0$ |
| • $\begin{cases} 1/\mu_4 \\ 0 \end{cases}$ | $0 < t < \mu_4$<br>$t > \mu_4 no$ |
| • $\begin{cases} \frac{1}{\mu_5}[1+\cos(\frac{\pi t}{\mu_5})] \\ 0 \end{cases}$ | $0 < t < \mu_5$<br>$t > \mu_5 no$ |
| • $\delta(t)$ | $t \geq 0$ |
| • any $g(t)$ \| Energy dissipation $> 0$ | $\int_0^\infty g(t)dt = 1,$<br>$t \geq 0$ |

In Table 1, $\mathbf{z_j^h}$ is the network input at layer $h$ and neuron $j$, $b_j$ is a scalar bias, and $\mathbf{s_i}$ is the output of the neuron $i$ at the layer $h-1$. Also, $v_{ij}$ is the kernel weight from the neuron $i$ at layer $h-1$ to the neuron $j$ at layer $h$, and $\mathbf{u_j^h}$ is the intermediate output. Related to the back propagation of the error starting from the output fully connected layer, $N_h$ is the number of classes in the input data, and $\mathbf{r_h}$ corresponds to the target and output vector. Finally, the delta of the neuron $j$ at layer $h$, $\Delta_h$, is used to update the bias of that neuron, as well as, all the weights of the neurons in the previous layer connected to that neuron.

## 6 APPLICATION TO LINEAR DYNAMIC SYSTEMS

For the linear numerical application consider the case of the damping model classes in structural dynamics. The standard equation of motion of a $n$ DOF structural-mechanical system, in the case of proportional damping, is written as:

$$\mathbf{M}\,\ddot{\mathbf{x}}(t) + \mathbf{C}\,\dot{\mathbf{x}}(t) + \mathbf{K}\,\mathbf{x}(t) = \mathbf{f}(t) \tag{16}$$

where, $\mathbf{M}$ and $\mathbf{K}$ are the mass and stiffness matrices, respectively, and $\mathbf{C}$ is the proportional to $\mathbf{M}$ and/or $\mathbf{K}$ damping matrix that satisfies the orthogonality property. This means that if $\mathbf{\Phi}$ is the matrix that contains the eigenvectors of the system, then $\overline{\mathbf{C}} = \mathbf{\Phi}^T\mathbf{C}\mathbf{\Phi}$ is a diagonal matrix and thus, a decoupling procedure can be implemented. Here, $\mathbf{x}(t)$ and $\mathbf{f}(t)$ are the response of the system and the force applied to the system, respectively.

With regard to damping the form of Equation (16) is restrictive, and for a general consideration of structural-mechanical systems, alternatively damping model classes are considered. This is implemented by one or more convolution integrals over a kernel function $g(t)$. In doing so, the damping depends on the past history of the motion. The equation of motion then is written as an integro-differential equation:

$$\mathbf{M}\,\ddot{\mathbf{x}}(t) + \mathbf{C}\int_0^t g(t-\tau)\,\dot{\mathbf{x}}(\tau)\,d\tau + \mathbf{K}\,\mathbf{x}(t) = \mathbf{f}(t) \tag{17}$$

where, this formulation is a generalization of the standard damping modeling since by using the Kronecker delta function $\delta(t)$ as the kernel function $g(t)$, Equation (17) reduces to Equation (16).

For the choice of the damping kernel functions, many candidate functions may be considered. Observations, though, from real systems[57] suggest that the exponential function can often adequately model the damping, and is a natural choice.

Table 2 shows several candidate kernel functions[13,58] which have been shown to adequately model the damping behavior of structural-mechanical systems. Here, $\mu_i$ is damping model parameter which is properly calibrated by system identification procedures.[13]

The system of Equation (17) is then examined with various model classes. Specifically, the system matrices for the synthetic measurement generation are:

$$\mathbf{M} = \begin{bmatrix} m_1 & 0 \\ 0 & m_2 \end{bmatrix} = \begin{bmatrix} 1 & 0 \\ 0 & 1 \end{bmatrix}, \quad \mathbf{C} = \begin{bmatrix} c_1 + c_2 & -c_2 \\ -c_2 & c_2 \end{bmatrix} = \begin{bmatrix} 1+2 & -2 \\ -2 & 2 \end{bmatrix},$$

$$\mathbf{K} = \begin{bmatrix} k_1 + k_2 & -k_2 \\ -k_2 & k_2 \end{bmatrix} = \begin{bmatrix} 9+11 & -11 \\ -11 & 11 \end{bmatrix} \tag{18}$$

with the initial conditions are $\mathbf{x}(0) = [1 \quad 1]^T$ and $\dot{\mathbf{x}}(0) = [0 \quad 0.5]^T$. White noise is chosen for the force $\mathbf{f}(t) = [f_1(t) \quad f_2(t)]^T$ of mean value 0 and variance 9. Importantly, the initial conditions and/or the force should be chosen to excite the system sufficiently.

Three model classes are considered with three different kernel functions, namely:

1. $g(t) = \mu_1 e^{-\mu_1 t}$ with $\mu_1 = 1.5$ (Model A)
2. $g(t) = 2\sqrt{\frac{\mu_3}{\pi}} e^{-\mu_3 t^2}$ with $\mu_3 = 1.5$ (Model B)
3. $g(t) = \delta(t)$ (Model C)

To create synthetic measurements, the integration method of Katsikadelis[13,59,60] is implemented as:

$$\boxed{\mathbf{z_k} = \mathbf{F_k} \cdot \mathbf{z_{k-1}} + \mathbf{B_k} \cdot \mathbf{u_k}} \tag{19}$$

where:

i) $\mathbf{z_{k-1}} = [\ddot{\mathbf{x}}_{\mathbf{k-1}} \quad \dot{\mathbf{x}}_{\mathbf{k-1}} \quad \mathbf{w_{k-1}} \quad \mathbf{x_{k-1}}]^T$

ii) $\mathbf{u_k} = [\mathbf{f_k} \quad \mathbf{0_{1\times n}} \quad \mathbf{0_{1\times n}} \quad \mathbf{f_{w_k}}]^T$

iii) $\mathbf{w_{k-1}} = \sum_{i=1}^{k-2} W_i \cdot (\dot{\mathbf{x}}_{\mathbf{i}} + \dot{\mathbf{x}}_{\mathbf{i-1}})/2 + W_{k-1} \cdot (\dot{\mathbf{x}}_{\mathbf{k-1}} + \dot{\mathbf{x}}_{\mathbf{k-2}})/2$

iv) $W_i = \int_{(i-1)\Delta t}^{(i)\Delta t} g(k \cdot \Delta t - \tau)\, d\tau$

v) $\mathbf{f_{w_k}} = \sum_{i=1}^{k-1} W_i \cdot (\dot{\mathbf{x}}_{\mathbf{i}} + \dot{\mathbf{x}}_{\mathbf{i-1}})/2$

vi) $$\mathbf{F}_k = \begin{bmatrix} \mathbf{M_{n\times n}} & \mathbf{0}_{n\times n} & \mathbf{C}_{n\times n} & \mathbf{K}_{n\times n} \\ \Delta t^2/4 \cdot \mathbf{I}_{n\times n} & -\Delta t \cdot \mathbf{I}_{n\times n} & \mathbf{0}_{n\times n} & \mathbf{I}_{n\times n} \\ -\Delta t/2 \cdot \mathbf{I}_{n\times n} & \mathbf{I}_{n\times n} & \mathbf{0}_{n\times n} & \mathbf{0}_{n\times n} \\ \mathbf{0}_{n\times n} & -W_k/2 \cdot \mathbf{I}_{n\times n} & \mathbf{I}_{n\times n} & \mathbf{0}_{n\times n} \end{bmatrix}^{-1} \begin{bmatrix} \mathbf{0_{n\times n}} & \mathbf{0}_{n\times n} & \mathbf{0}_{n\times n} & \mathbf{0}_{n\times n} \\ -\Delta t^2/4 \cdot \mathbf{I}_{n\times n} & \mathbf{0}_{n\times n} & \mathbf{0}_{n\times n} & \mathbf{I}_{n\times n} \\ \Delta t/2 \cdot \mathbf{I}_{n\times n} & \mathbf{I}_{n\times n} & \mathbf{0}_{n\times n} & \mathbf{0}_{n\times n} \\ \mathbf{0}_{n\times n} & W_k/2 \cdot \mathbf{I}_{n\times n} & \mathbf{0}_{n\times n} & \mathbf{0}_{n\times n} \end{bmatrix}$$

vii) $$\mathbf{B}_k = \begin{bmatrix} \mathbf{M_{n\times n}} & \mathbf{0}_{n\times n} & \mathbf{C}_{n\times n} & \mathbf{K}_{n\times n} \\ \Delta t^2/4 \cdot \mathbf{I}_{n\times n} & -\Delta t \cdot \mathbf{I}_{n\times n} & \mathbf{0}_{n\times n} & \mathbf{I}_{n\times n} \\ -\Delta t/2 \cdot \mathbf{I}_{n\times n} & \mathbf{I}_{n\times n} & \mathbf{0}_{n\times n} & \mathbf{0}_{n\times n} \\ \mathbf{0}_{n\times n} & -W_k/2 \cdot \mathbf{I}_{n\times n} & \mathbf{I}_{n\times n} & \mathbf{0}_{n\times n} \end{bmatrix}^{-1}$$

Here, the time discretization frequency is set equal to 100 Hz, therefore $\Delta t$ is 0.01. The same holds for the sampling frequency of the synthetic measurements. Finally, to consider the effect of measurement noise, each response signal is contaminated by a Gaussian white noise sequence with a 10% root-mean-square noise-to-signal ratio. Different initial conditions are applied to the system to generate multiple responses for training and validation. The duration of the acceleration and displacement signal measurement for each model class is 40 s.

To Kalman filter all previous signals, the process covariance $\mathbf{Q_d}$ and the measurement covariance $\mathbf{R_d}$ matrices are chosen to be constant during the identification process and equal to $10^{-9} \cdot \mathbf{I_{2\times 2}}$ and $10^{-3} \cdot \mathbf{I_{1\times 1}}$, respectively. For larger values, the algorithm needs more data and time to converge, or it may even diverge.

The convolutional neural network architecture is defined as follows in Figure 1: An input layer with the three signals for each one of the three model classes A, B, and C, associated with their model class label. A convolutional layer is set with filter size equal to 2048 and number of neurons that connect to the same region of the input equal to 128 with casual padding. A rectifier layer, termed also as ReLu is also set, as well as a batch normalization layer with mini-Batch size equal to 1 for online purposes, and an additional convolutional layer with filter size equal to 2048 and number of neurons that connect to the same region of the input equal to 256 with casual padding. An additional rectifier layer is set along with an additional batch normalization layer with mini-Batch size equal to 1, and a global average pooling layer. Finally, a fully connected layer is set with a number of classes equal to 3, a softmax layer, and a classification layer. Importantly, an investigation of the number of the filter size and the number of neurons within the convolutional layer is shown in Section 9. Last but not least, the number of the maximum epochs in the optimization process is set equal to 15. Importantly, to design the architecture, although still an active research problem,[61] a simple CNN architecture is examined that has one hidden layer with one max pooling layer before the classification one. Based on the results and by controlling the trade-off between accuracy and training speed, the number of kernels and layers is increased until a satisfactory performance is reached. This work uses similar architecture building philosophy to the damage detection applications[28,32] without any special adjustments that would potentially favor the model class selection problem.

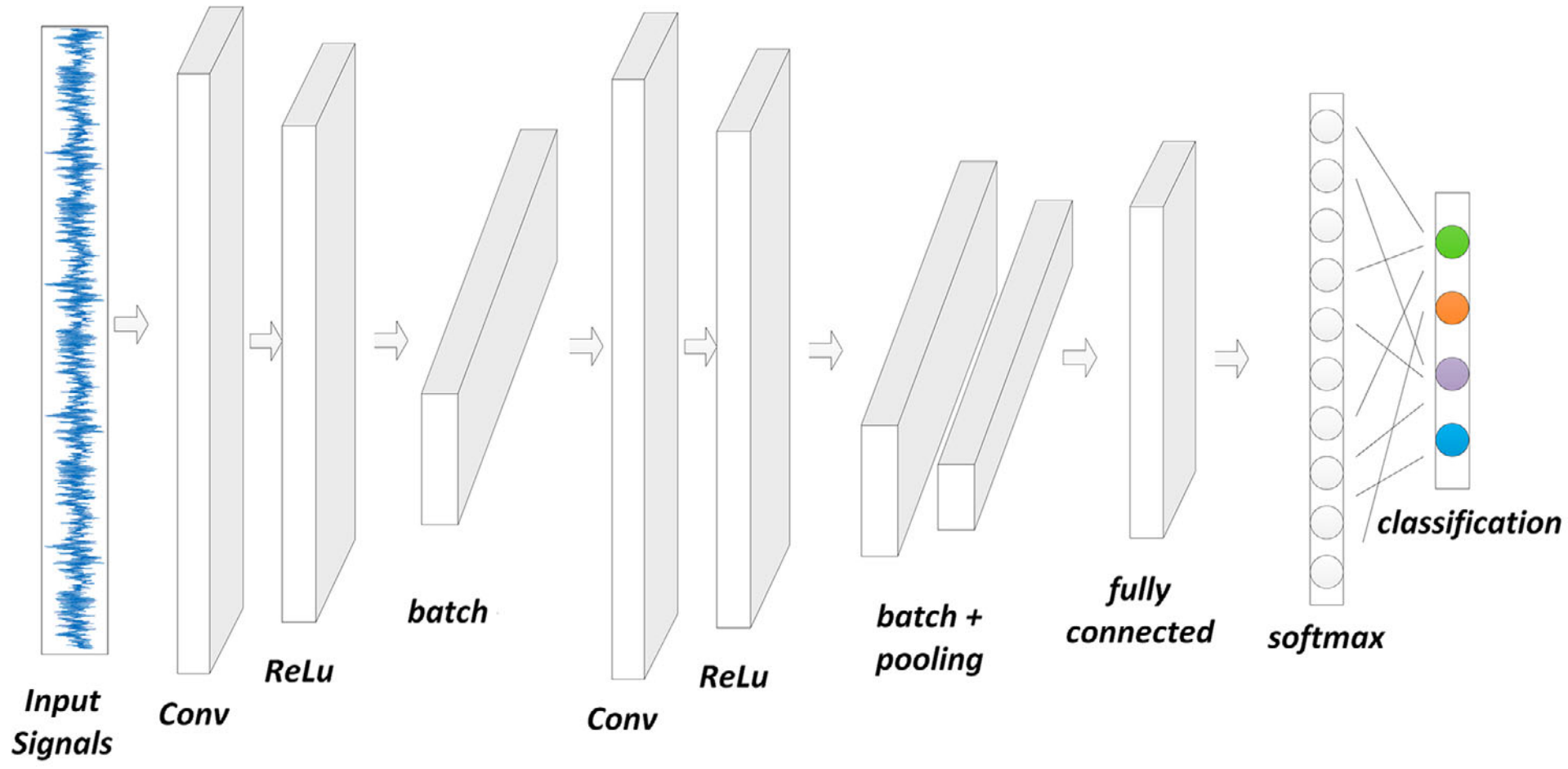

**FIGURE 1** Examined C-Net architecture for all numerical applications.

Two signal inputs are examined in Figures 2–3. In these figures, the first and second row refer to the displacement and acceleration raw signal used in the Kalman filter for all models. The third row refers to the network model class selection trained with unfiltered signals (C-Net) where the data generated by model A denoted by 1, model B denoted by 2, and model C denoted by 3 are attributed to each model A, B, or C. Similarly, the fourth row refers to the network model class selection trained with the Kalman filtered signals (Kalman filter C-Net). Additionally, the fifth row refers to the accuracy in the training process for both networks with respect to the number of optimization iterations, while the sixth row refers to the loss in the training process for both networks with respect to the number of optimization iterations. In total, 9 new velocity and displacement signals are classified, where ideally the first 3 signals belong to Model A, the second 3 signals belong to Model B, and the last 3 signals belong to Model C.

In Figure 2, the performance of the networks using only the DOF 2 displacement signals in the training and validation process is shown. The C-Net correctly selects the model class for each signal. The Kalman filter C-Net also correctly selects the model class for each signal, but with a shorter training period and loss minimization than C-Net.

In Figure 3, the performance of the networks using only the DOF 2 velocity signals in the training and validation process is shown. Both networks select correctly the model class for each signal expect one. Importantly, the Kalman filter C-Net converges faster.

Importantly, in this application, it may seem that the model is quite simple and, perhaps, does not need such a complex network in the prediction, meaning the convolution neural network is not efficiently designed. In reality though, removing layers from the network results in a poorer performance where the predictions are wrong.

Additionally, it may seem that the model classes are too idealized since the model class can be well depicted by the mathematical formulas in Table 2. In reality though, those models have been experimentally demonstrated that they represent the behavior of real dynamic systems, that is, chap. 8 of Adhikari.[57]

## 7 | APPLICATION TO NONLINEAR DYNAMIC SYSTEMS

For the nonlinear numerical application consider initially the problem of a mass in free fall[62] landing on a generalized damped base material. The stiffness and damping elements of the base material are active only when the body is in contact

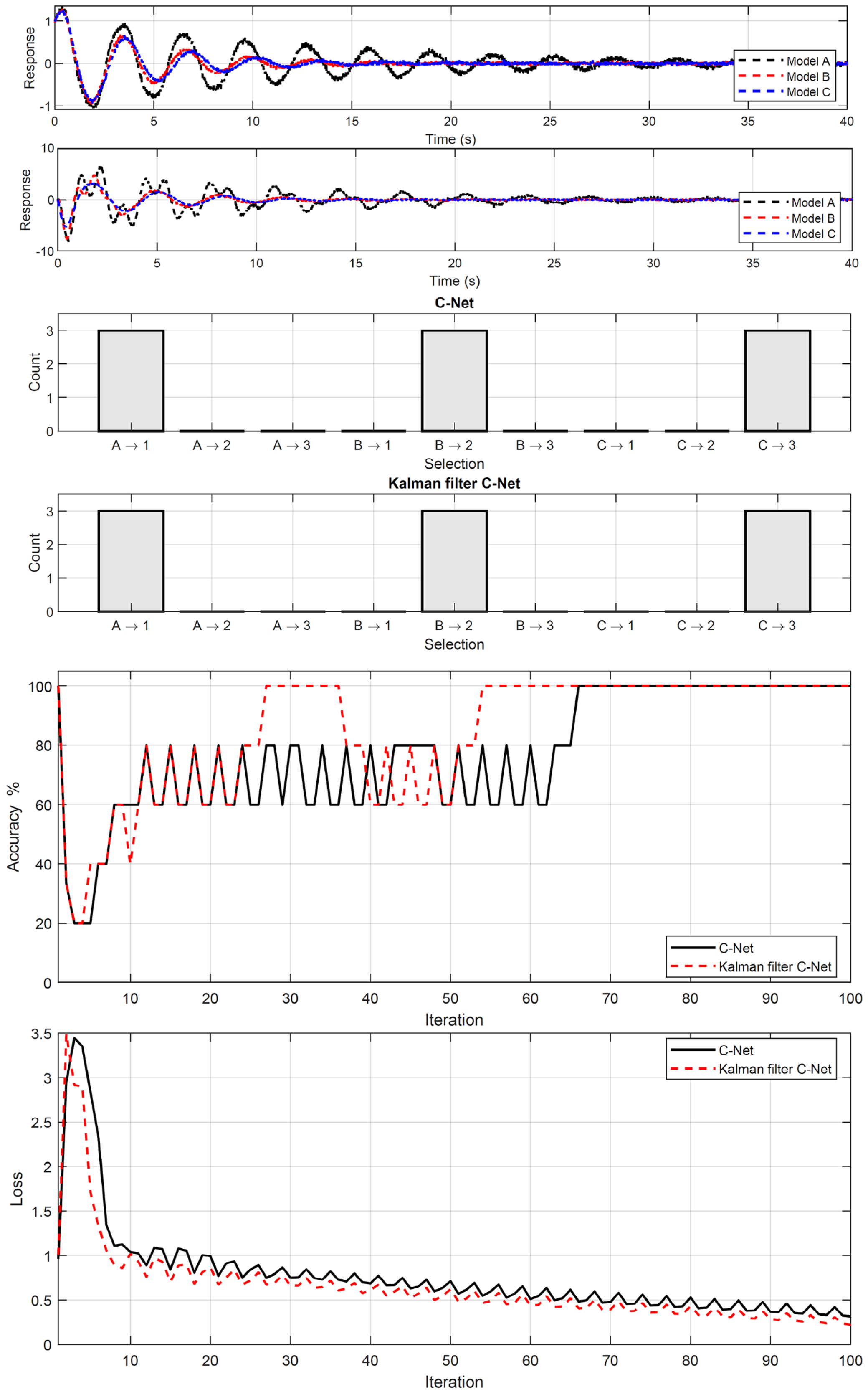

**FIGURE 2** System of Section 6: Results for the linear dynamic system when training and validating with the DOF 2 displacement signals. First and second row: the displacement and acceleration raw signals in m/s and m/s$^2$, respectively. Third row: C-Net model class prediction where ideally A->1, B->2, and C->3. Fourth row: Kalman filter C-Net model class prediction. Fifth and six row: accuracy and loss in the training process for both networks.

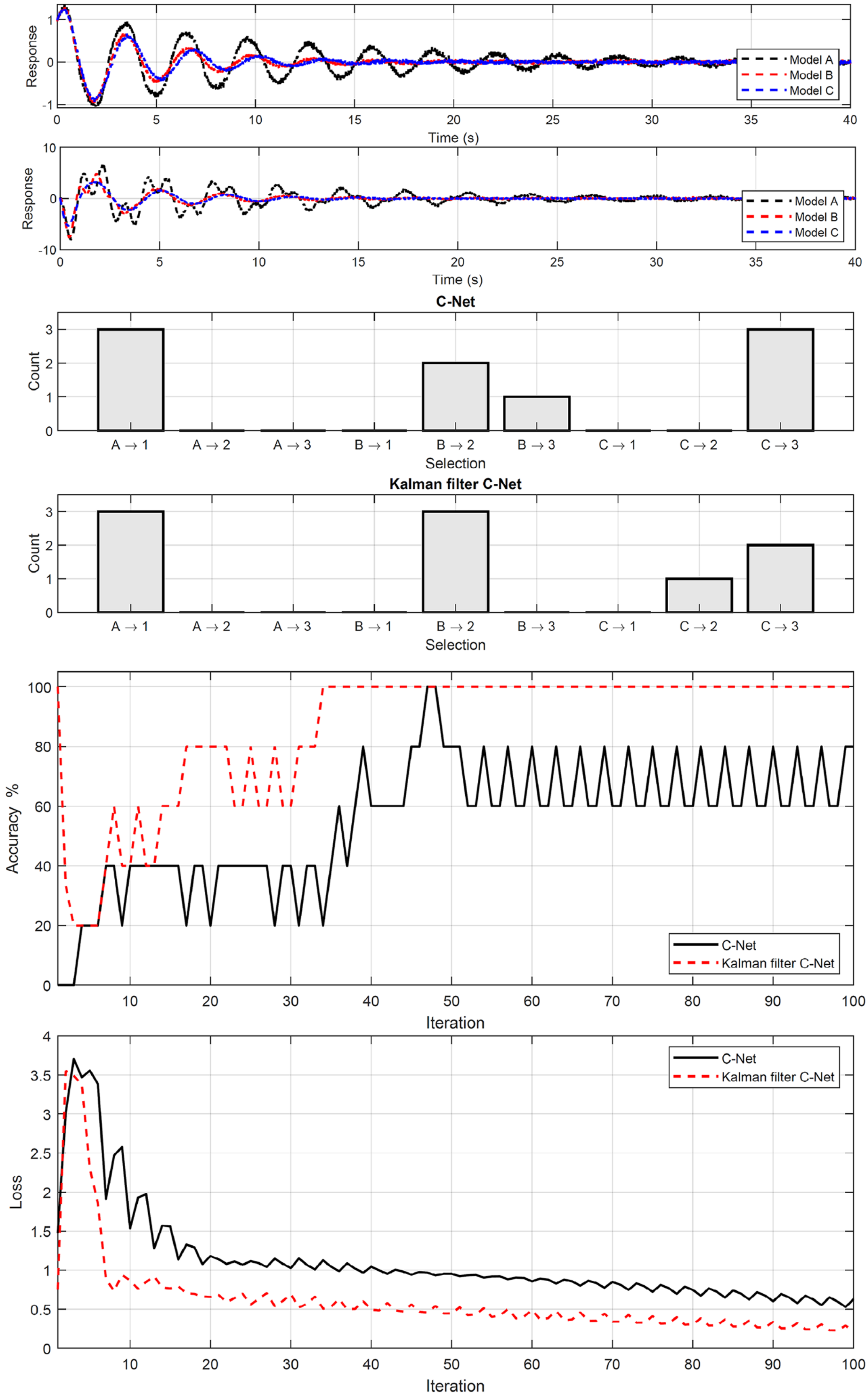

**FIGURE 3** System of Section 6: Results for the linear dynamic system when training and validating with the DOF 2 velocity signals. First and second row: the displacement and acceleration raw signals in m/s and m/s$^2$, respectively. Third row: C-Net model class prediction where ideally A->1, B->2, and C->3. Fourth row: Kalman filter C-Net model class prediction. Fifth and six row: accuracy and loss in the training process for both networks. DOF, degree of freedom.

with it. The equation of motion is nonlinear and it is expressed as:

$$m\,\ddot{x}(t) + G(x(t),\,\theta) = f(t) \tag{20}$$

and if, for instance, the effectiveness of a twofold model is examined, then the equation of motion is written as:

$$m\,\ddot{x}(t) + \mathbb{H}(x(t)) \cdot \left\{ c \cdot \int_0^t \delta(t-\tau)\,\dot{x}(\tau)\,d\tau + c \cdot \int_0^t g(t-\tau)\,\dot{x}(\tau)\,d\tau + k\,x(t) \right\} = f(t) \tag{21}$$

where, $\mathbb{H}(x(t))$ is the Heaviside step function. Assume here, $m = 1$ Kg, $c = 3$ N s/m, $k = 1000$ N/m, $f(t) = -mg$ with $g = 9.81\ m/s^2$ for the gravity acceleration. The initial conditions are $\mathbf{x}(0) = 0.1$ and $\dot{\mathbf{x}}(0) = 0$. White noise is chosen for the force $\mathbf{f}(t)$ of mean value 0 and variance 9. Importantly, the initial conditions and/or the force should be chosen to excite the system sufficiently.

Two model classes are considered for the $g(t)$ in Equation (21) with two different kernel functions, namely:

1. $g(t) = \mu_1 e^{-\mu_1 t}$ with $\mu_1 = 100$ (Model A)
2. $g(t) = 2\sqrt{\frac{\mu_3}{\pi}} e^{-\mu_3 t^2}$ with $\mu_3 = 100$ (Model B)

To create synthetic measurements, the integration method of Katsikadelis[13,59,60] is implemented as in Section 6 where, for nonlinear systems, the state transition matrix $\mathbf{F_k}$ and the input matrix $\mathbf{B_k}$ are modified and in the stead of $\mathbf{C}_{n\times n}$ and $\mathbf{K}_{n\times n}$, the zero matrix $\mathbf{0}_{n\times n}$ is inserted. Also, the new input is:

$$\mathbf{u_k} = [(\mathbf{f_k} - \mathbf{f_{n_k}}) \quad \mathbf{0_{1\times n}} \quad \mathbf{0_{1\times n}} \quad \mathbf{f_{w_k}}]^T \tag{22}$$

where, a system of equations provides the numerical solution of the nonlinear system, namely:

$$\begin{cases} \mathbf{z_k} = \mathbf{F_k} \cdot \mathbf{z_{k-1}} + \mathbf{B_k} \cdot \mathbf{u_k} \\ \mathbf{f_{n_k}} = \mathbf{G}(\mathbf{z_k},\, \boldsymbol{\theta_k}) \end{cases} \tag{23}$$

Here, the time discretization frequency is set equal to 100 Hz, therefore $\Delta t$ is 0.01. The same holds for the sampling frequency of the synthetic measurements. Finally, to consider the effect of measurement noise, each response signal is contaminated by a Gaussian white noise sequence with a 10% root-mean-square noise-to-signal ratio. The duration of the acceleration and displacement signal measurement for each model class is 100 s.

To Kalman filter the signals, the process covariance $\mathbf{Q_d}$ and the measurement covariance $\mathbf{R_d}$ matrices are chosen to be constant during the identification process and equal to $10^{-9} \cdot \mathbf{I_{2\times 2}}$ and $10^{-3} \cdot \mathbf{I_{1\times 1}}$, respectively. For larger values, the algorithm needs more data and time to converge, or it may even diverge.

Subsequently, the network architecture is defined similarly to Section 6.

Two signal inputs are examined in Figures 4–5 with the same layout description as in Section 6. In total, 10 new velocity and displacement signals are classified, where ideally the first 5 signals belong to Model A, and the second 5 signals belong to Model B.

In Figure 4, the performance of the networks using only the DOF 1 displacement signals in the training and validation process is shown. The C-Net correctly selects the model class for each signal apart from one which is misclassified as Model A despite belonging to Model B. The Kalman filter C-Net also provides the same selection accuracy, but with a shorter training period and loss minimization.

In Figure 5, the performance of the networks using only the DOF 1 velocity signals in the training and validation process is shown. The C-Net selects correctly the class of seven signals, but misselects three of them. Contrastingly, the Kalman filter C-Net misselects only 1 signal out 10. In this examination, the Kalman filter C-Net shows a superior performance compared to C-Net in the selection accuracy, apart from solely a faster convergence.

Furthermore, for the nonlinear numerical application to other model class types and not only in damping kernels such as on the stiffness matrix,[63] consider a 6-story shear type model extending the application of Kontoroupi and Smyth.[12,47] Here, the first DOF is associated with a nonlinear hysteretic behavior based on the Bouc–Wen model[64] which has shown a great practical potential in structural engineering.[65–78] Three candidate models are considered here which differ in the expression of the hysteretic component $r_1$, namely:

1. $\dot{r}_1 = A\,\dot{x}_1 - (\beta\,|\dot{x}_1|\,r_1^{n-1}\,r_1 + \gamma\,\dot{x}_1\,|r_1|^n)$ (Model A - without degradation[64])
2. $\dot{r}_1 = \dfrac{A\,\dot{x}_1 - (\beta\,|\dot{x}_1|\,r_1^{n-1}\,r_1 + \gamma\,\dot{x}_1\,|r_1|^n)}{\eta(t)}$ with $\dot{\epsilon}_1 = r_1\,\dot{x}_1$, $\eta(t) = 1 + \delta_\eta\,\epsilon_1(t)$ (Model B - with degradation[79])
3. $\dot{r}_1 = \dfrac{1 - (\beta\,sing(\dot{x}_1)\,|r_1|^{n-1}\,r_1 + \gamma\,|r_1|^n)}{1 + \sqrt{\dfrac{2}{\pi}}\,\dfrac{s(t)}{\sigma}\,exp[-\dfrac{r_1^2}{2\,\sigma^2}]\,[1 - (\beta\,sing(\dot{x}_1)\,|r_1|^{n-1}\,r_1 + \gamma\,|r_1|^n)]}$ with $s(t) = \delta_\sigma\,\epsilon_1(t)$ (Model C - with pinching[80])

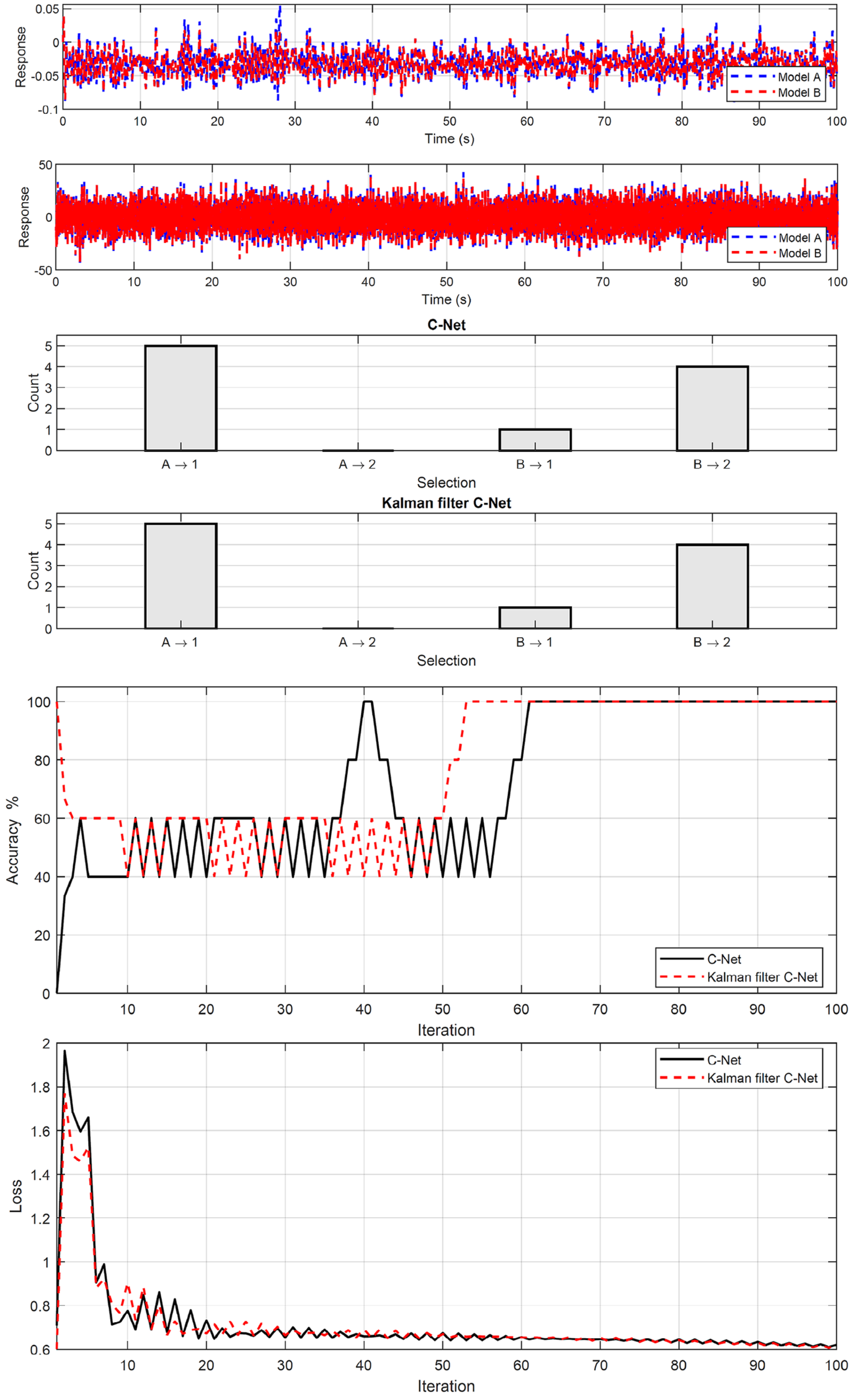

**FIGURE 4** System of Section 7: Results for the free fall nonlinear system when training and validating with the DOF 1 displacement signals. First and second row: the displacement and acceleration raw signals in m/s and m/s$^2$, respectively. Third row: C-Net model class prediction where ideally A->1 and B->2. Fourth row: Kalman filter C-Net model class prediction. Fifth and six row: accuracy and loss in the training process for both networks. DOF, degree of freedom.

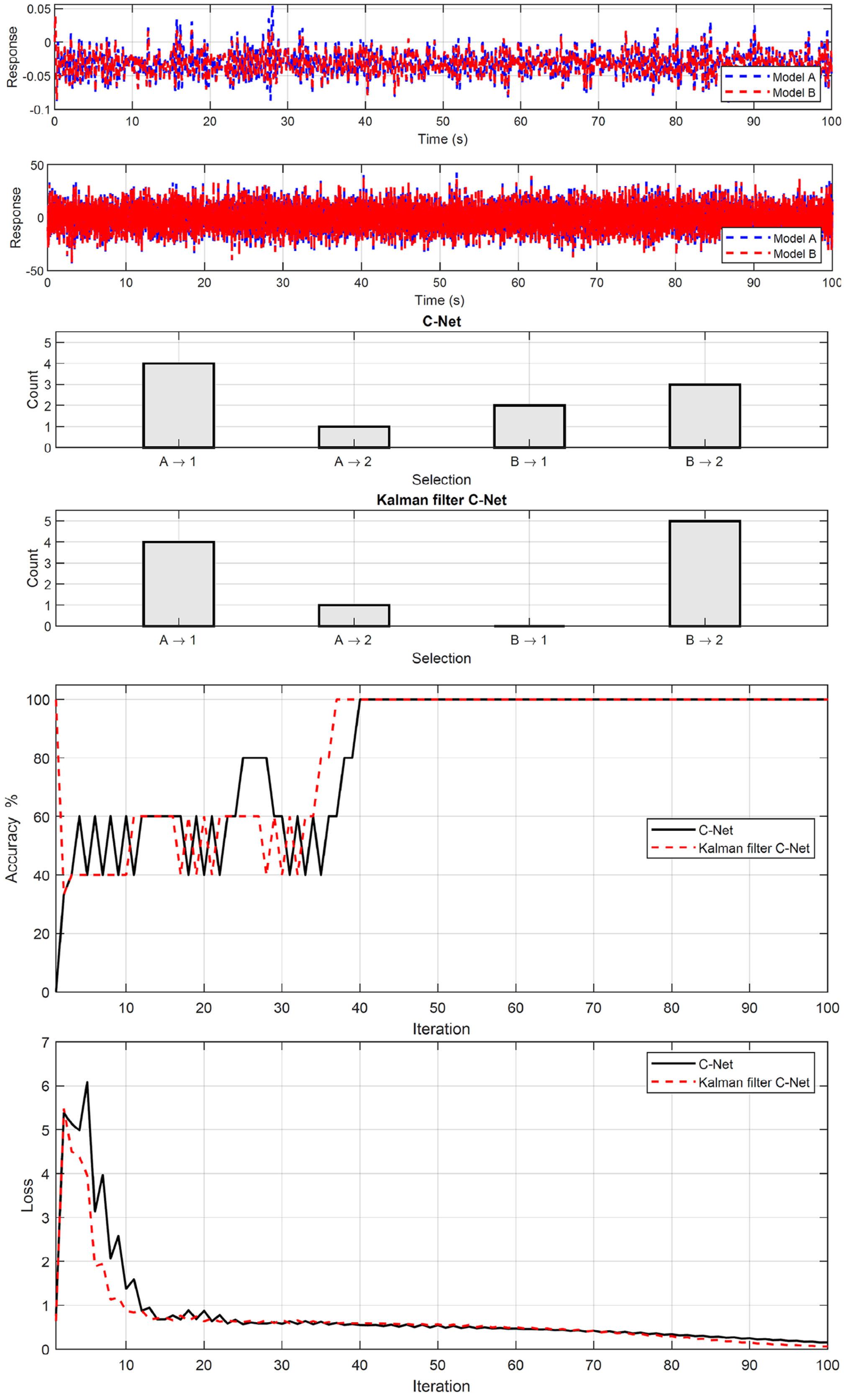

**FIGURE 5** System of Section 7: Results for the free fall nonlinear system when training and validating with the DOF 1 velocity signals. First and second row: the displacement and acceleration raw signals in m/s and m/s$^2$, respectively. Third row: C-Net model class prediction where ideally A->1 and B->2. Fourth row: Kalman filter C-Net model class prediction. Fifth and six row: accuracy and loss in the training process for both networks. DOF, degree of freedom.

To create synthetic measurements, the fourth-order Runge–Kutta integration method is used with $m_i = 1$, $k_i = 9$, $c_i = 0.25$, $A = 1$, $\beta = 2$, $\gamma = 1$, $n = 2$, $\delta_\eta = 0.4$, $\sigma = 0.1$, and $\delta_\sigma = 0.4$. Here, the time discretization frequency is set equal to 50 Hz, therefore $\Delta t$ is 0.02. The same holds for the sampling frequency of the synthetic measurements. Finally, to consider the effect of measurement noise, each response signal is contaminated by a Gaussian white noise sequence with a 10% root-mean-square noise-to-signal ratio.

To train the network, three earthquake inputs are considered, namely the Tabas of September 16, 1978 at Tabas (1.080 g), the Northridge of January 17, 1994 at Sylmar Converter Station (0.827 g), and the Kobe of January 17, 1995 at JMA (0.818 g), available from the Sylmar Converter Station (PEER strong motion database[81]). Only those three are used for training the convolutional neural network, while three more are used for the validation step.

To Kalman filter the signals, the process covariance $\mathbf{Q_d}$ and the measurement covariance $\mathbf{R_d}$ matrices are chosen to be constant during the identification process and equal to $10^{-9} \cdot \mathbf{I_{2\times2}}$ and $10^{-3} \cdot \mathbf{I_{1\times1}}$, respectively. For larger values, the algorithm needs more data and time to converge, or it may even diverge.

Subsequently, the network architecture is defined similarly to Section 6.

Two signal inputs are examined in Figures 6–7 with the same layout description as in Section 6. In total, 9 new velocity and displacement signals are classified, where ideally the first 3 signals belong to Model A, the second 3 signals belong to Model B, and the final 3 signals belong to Model C.

In Figure 6, the performance of the networks using only the DOF 1 displacement signals in the training and validation process is shown. The C-Net correctly selects the model class for each signal apart from one which is misclassified as Model A despite belonging to Model B. The Kalman filter C-Net also provides the same selection accuracy, but with a shorter training period and loss minimization.

In Figure 7, the performance of the networks using only the DOF 1 velocity signals in the training and validation process is shown. The C-Net selects correctly the class of eight signals, but misselects one of them. The Kalman filter C-Net also provides the same selection accuracy, but with a shorter training period and loss minimization.

## 8 | APPLICATION TO A 3D BUILDING FINITE ELEMENT MODEL

For the 3D building finite element model application consider the problem of the N-storey building of Figure 8 simulated in OpenSees,[82] which has show a great potential for capturing the realistic behavior of structures.[83–86] This problem examines the capability of the approach when due to the large number of DOFs, the network may not capture all the dynamic system changes and become inaccurate.

The model has six DOFs at each node of a studied 2-storey and 2-bay at each direction 3D model. Each column has a length of 14 feet (4.3 m) with section W27x114, each beam has a length of 24 feet (7.3 m) with section W24x94, and each girder has a length of 24 feet with section W24x94. The ground boundary are assumed fixed, and the material properties are 29,000 Ksi (200 GPa) for the Elastic modulus, 0.3 for the Poisson ratio, and 60 Ksi (413.6 MPa) for the yield stress. A hardening material law is chosen.[87] The weight of all components is taken into account, and reinforced-concrete floor slabs are simulated with 150 pcf (2403 Kg/m$^3$) concrete density and scale factor 2 for dead loads. Importantly, the forceBeamColumn element is used for all components.[88]

Two model classes are considered for the Rayleigh damping[89] proportional to the matrix (Model A where, $\mathbf{C} = \alpha_1\,\mathbf{M}$), or proportional to both the mass and the stiffness matrix (Model B where, $\mathbf{C} = \alpha_1\,\mathbf{M} + \alpha_2\,\mathbf{K}$) with the Reyleigh damping parameters $\alpha_1$ and $\alpha_2$.

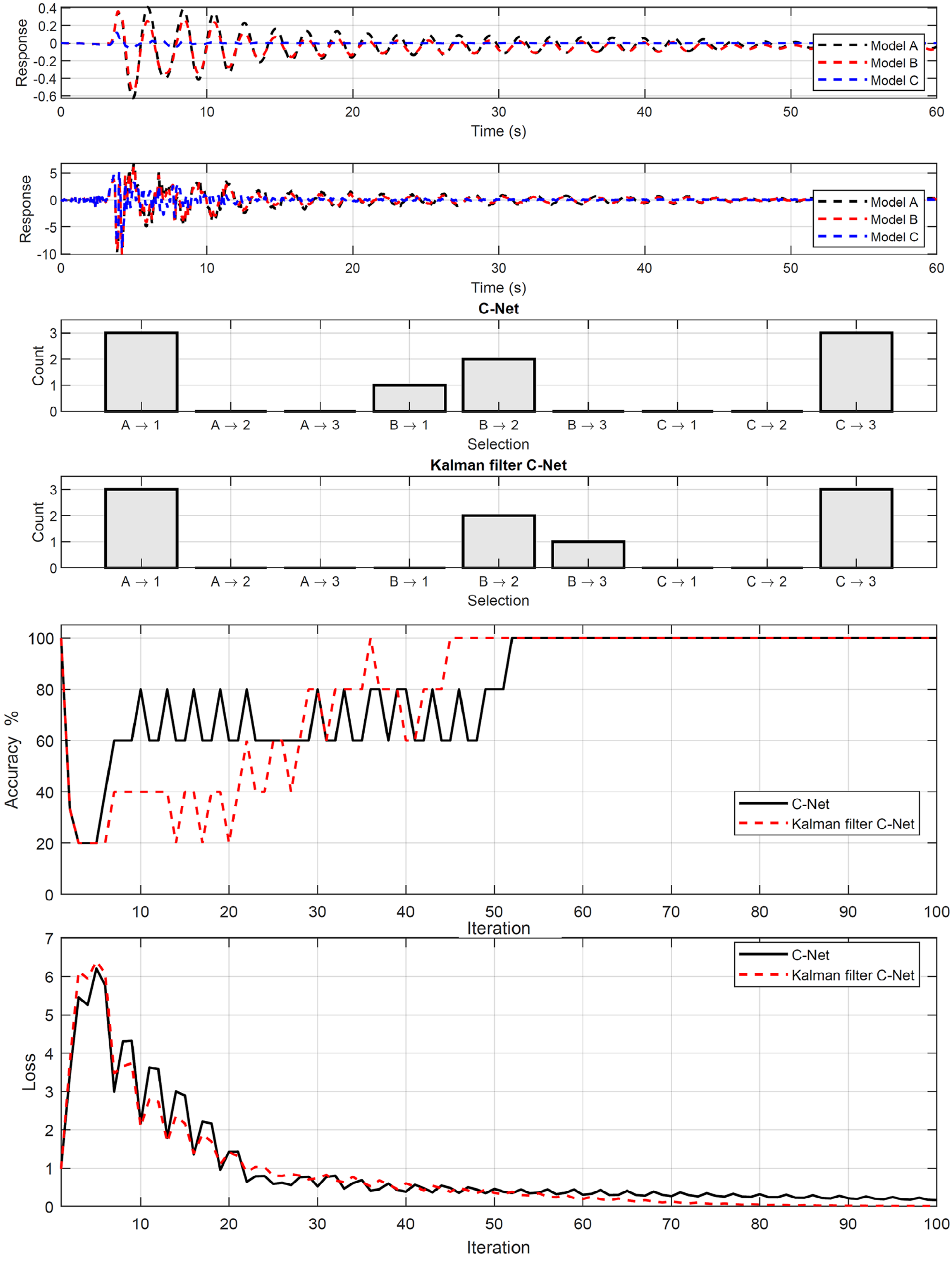

**FIGURE 6** System of Section 7: Results for the hysteretic nonlinear system when training and validating with the DOF 1 displacement signals. First and second row: the displacement and acceleration raw signals in m/s and m/s$^2$, respectively. Third row: C-Net model class prediction where ideally A->1, B->2, and C->3. Fourth row: Kalman filter C-Net model class prediction. Fifth and six row: accuracy and loss in the training process for both networks. DOF, degree of freedom.

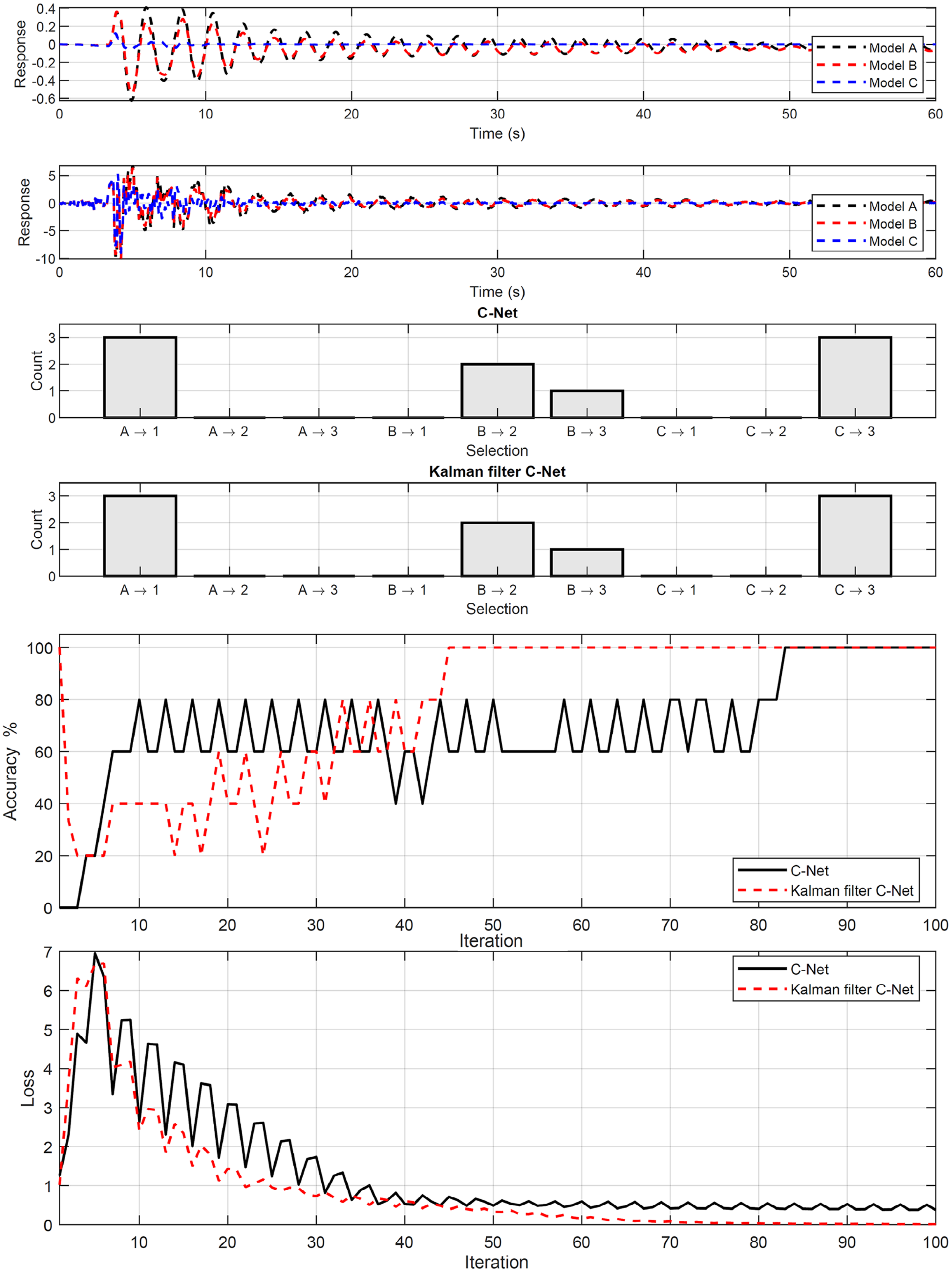

**FIGURE 7** System of Section 7: Results for the hysteretic nonlinear system when training and validating with the DOF 1 velocity signals. First and second row: the displacement and acceleration raw signals in m/s and m/s$^2$, respectively. Third row: C-Net model class prediction where ideally A->1, B->2, and C->3 Fourth row: Kalman filter C-Net model class prediction. Fifth and six row: accuracy and loss in the training process for both networks. DOF, degree of freedom.

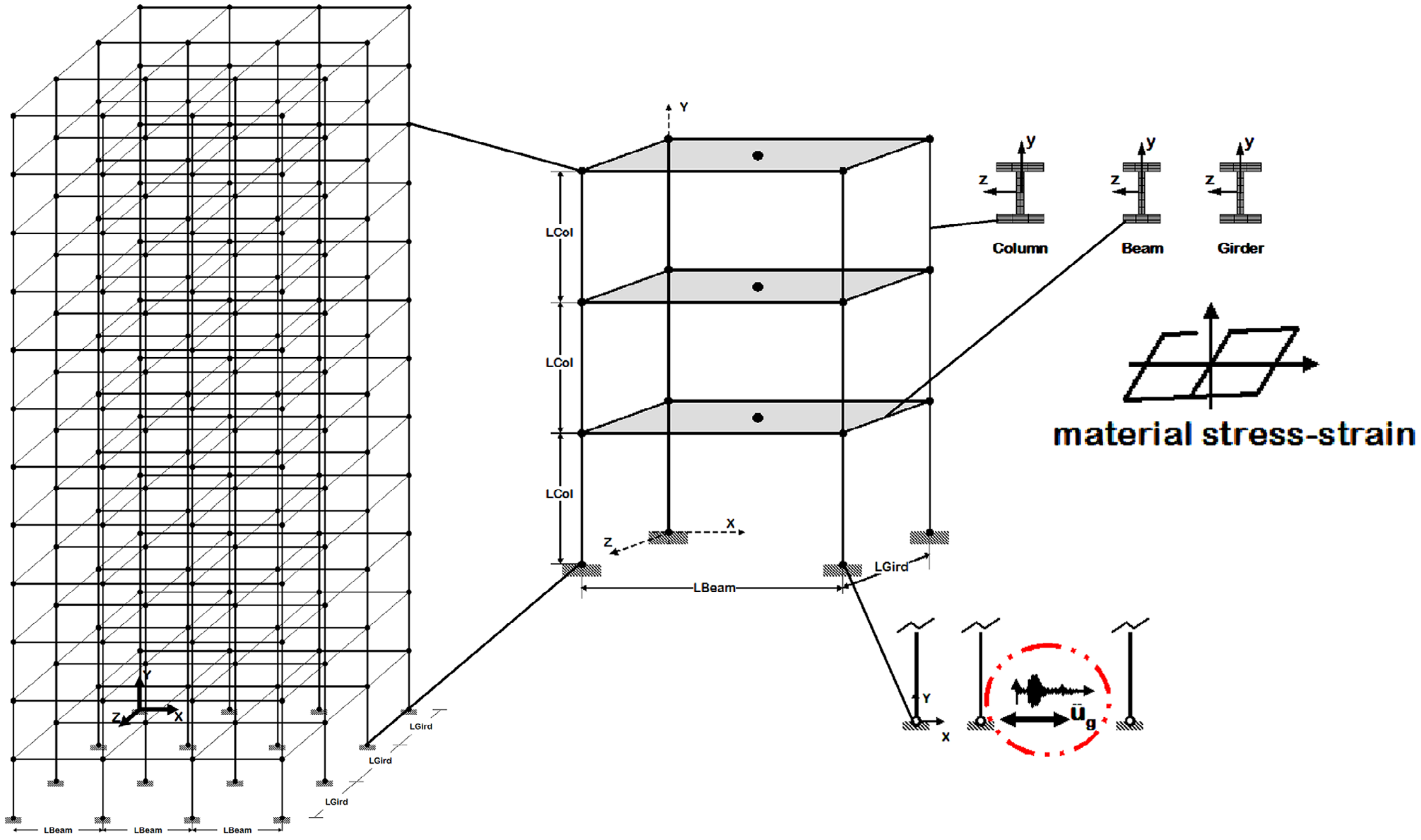

**FIGURE 8** 3D building finite element model system of Section 8 with material nonlinearity excited by earthquake inputs for the nonlinear history response calculation using OpenSees. DOF, degree of freedom.

To create synthetic measurements, the Newmark integration method is used to simulate the response with either the Newton with initial tangent or the Newton with line search method for the material nonlinearity, depending on the convergence issues.

Here, the time discretization frequency is set equal to 50 Hz, therefore $\Delta t$ is 0.02. The same holds for the sampling frequency of the response measurements. Finally, to consider the effect of measurement noise, each response signal is contaminated by a Gaussian white noise sequence with a 10% root-mean-square noise-to-signal ratio.

To train the network, three earthquake inputs are considered, namely, the Imperial Valley of May 18, 1940 at El Centro (0.341 g), the Northridge of January 17, 1994 at Sylmar Converter Station (0.827 g), and the Kobe of January 17, 1995 at JMA (0.818 g), available from the Sylmar Converter Station (PEER strong motion database[81]).

Only those three responses are used for training the convolutional neural network, while three more are used for the validation step. In this application, it is shown solely the C-Net performance to compare the training with acceleration signals which are not available in a filtered fashion by this Kalman filter approach. Importantly, to better illustrate the feasibility of the research in real buildings, the seismic responses of model is usually compared with some deformation index, such as story drift ratio, which can represent the deformation state of the structure. In the examined application this range is 0%–2%. However, in this work is not reported in detail to follow the unique DOF measurement approach for model class selection as examined earlier. The network architecture is defined similarly to Section 6.

Three signal inputs are examined in Figures 9–11 with similar layout description as in Section 6. In total, 10 new displacement, velocity, and acceleration signals are classified, where ideally the first 5 signals belong to Model A, and the second 5 signals belong to Model B.

In Figure 9, the performance of the network is shown using only the top corner building DOF displacement signals. The C-Net correctly selects the model class for each signal apart from one which is misclassified as Model A despite belonging to Model B, and one which is misselected as Model B although belonging to Model A.

In Figure 10, the performance of the network is shown using only the top corner building DOF velocity signals. The C-Net correctly selects the model class for each signal apart from one which is misclassified as Model A despite belonging to Model B.

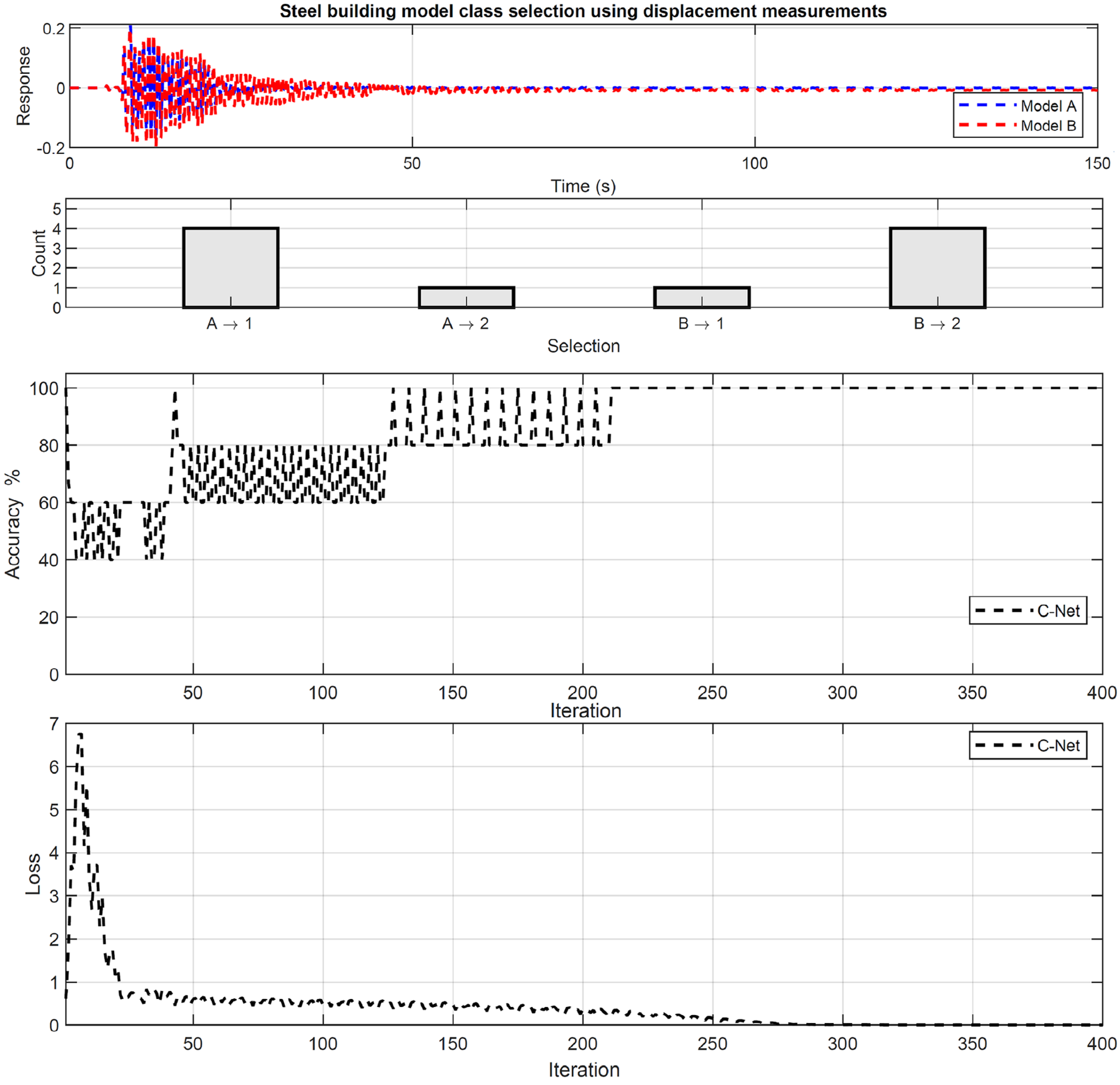

**FIGURE 9** System of Section 8: Results for the 3D building finite element model when training and validating with the top corner DOF displacement signals (Kobe plot). First row: the displacement raw signals in *m*. Second row: C-Net model class prediction where ideally A->1 and B->2. Third and fourth row: accuracy and loss in the training process. DOF, degree of freedom.

Finally, in Figure 11, the performance of the network is shown using only the top corner building DOF acceleration signals. The C-Net correctly selects the model class for each signal apart from two which are misclassified.

## 9 | DISCUSSION

The presented work provided a simple, yet effective, way to select the model class in structural dynamics. It did not aim to present a machine learning algorithm advancement, rather than to apply the vast capabilities of such tools[90–95] to the model class selection problem, for the first time to the best of the author's knowledge. To this end, the efficiency and robustness of the method was tested to both low-DOF systems and to a complex system, such as a 3D building finite element model. Further examinations and comparisons are also provided in the section to shed light into the method.

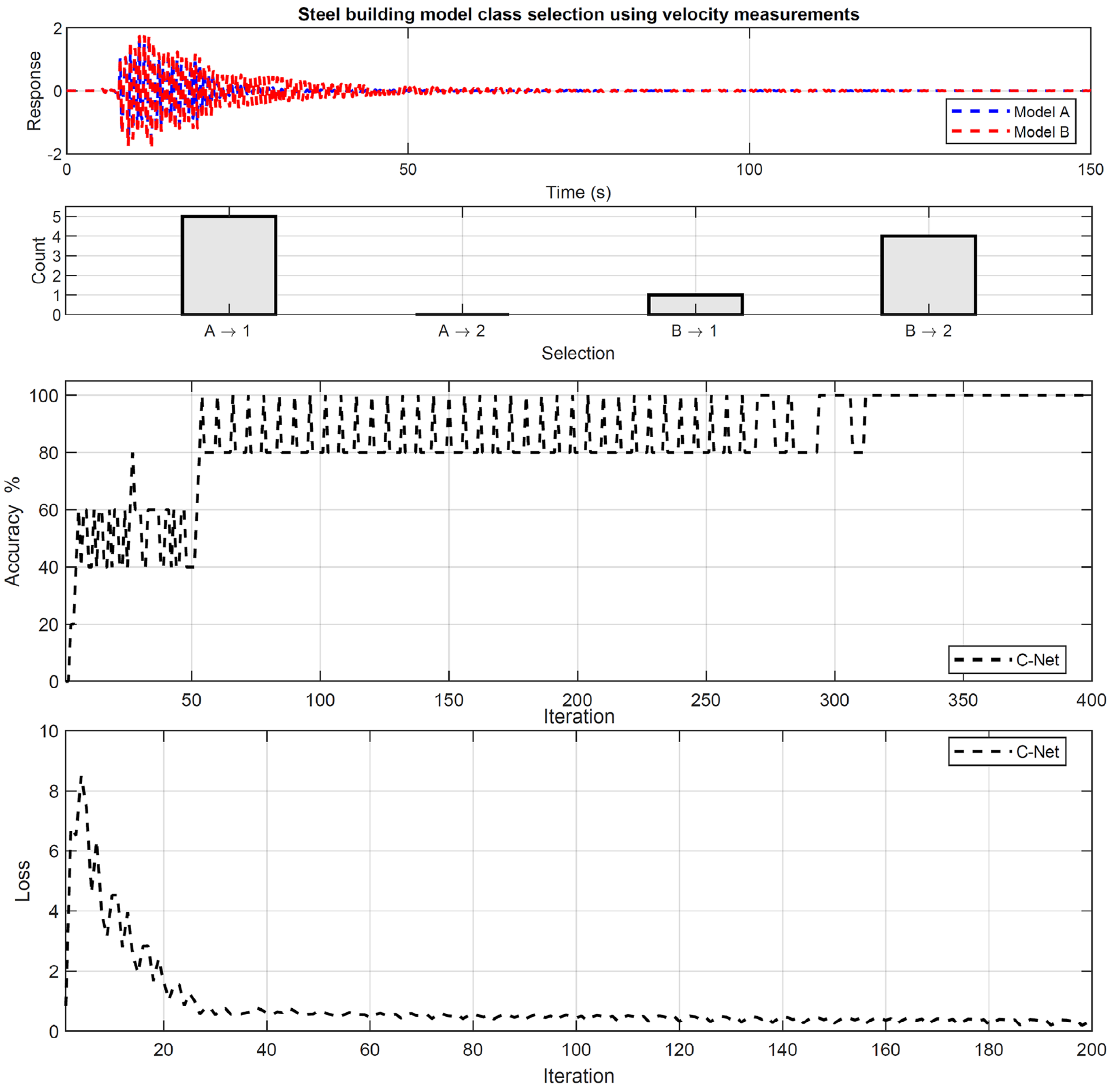

**FIGURE 10** System of Section 8: Results for the 3D building finite element model when training and validating with the top corner DOF velocity signals (Kobe plot). First row: the displacement raw signals in m/s. Second row: C-Net model class prediction where ideally A->1 and B->2. Third and fourth row: accuracy and loss in the training process. DOF, degree of freedom.

Specifically, the comparison between C-Net and Kalman filter C-Net may seem not fair. In the Kalman filter C-Net, the availability of the dynamic states provides more information compared to pure C-net, and this leads to a better accuracy since it has deeper information. In reality, the purpose of this work is not to improve the C-Net, but to provide a way to exploit more data if available. Importantly for the explanation of the results, the Kalman filter approach provides improved training performance since it exploits the estimated dynamic states which have less noise; however this impact is irrelevant when poor filter size and number of neurons is used for the network.

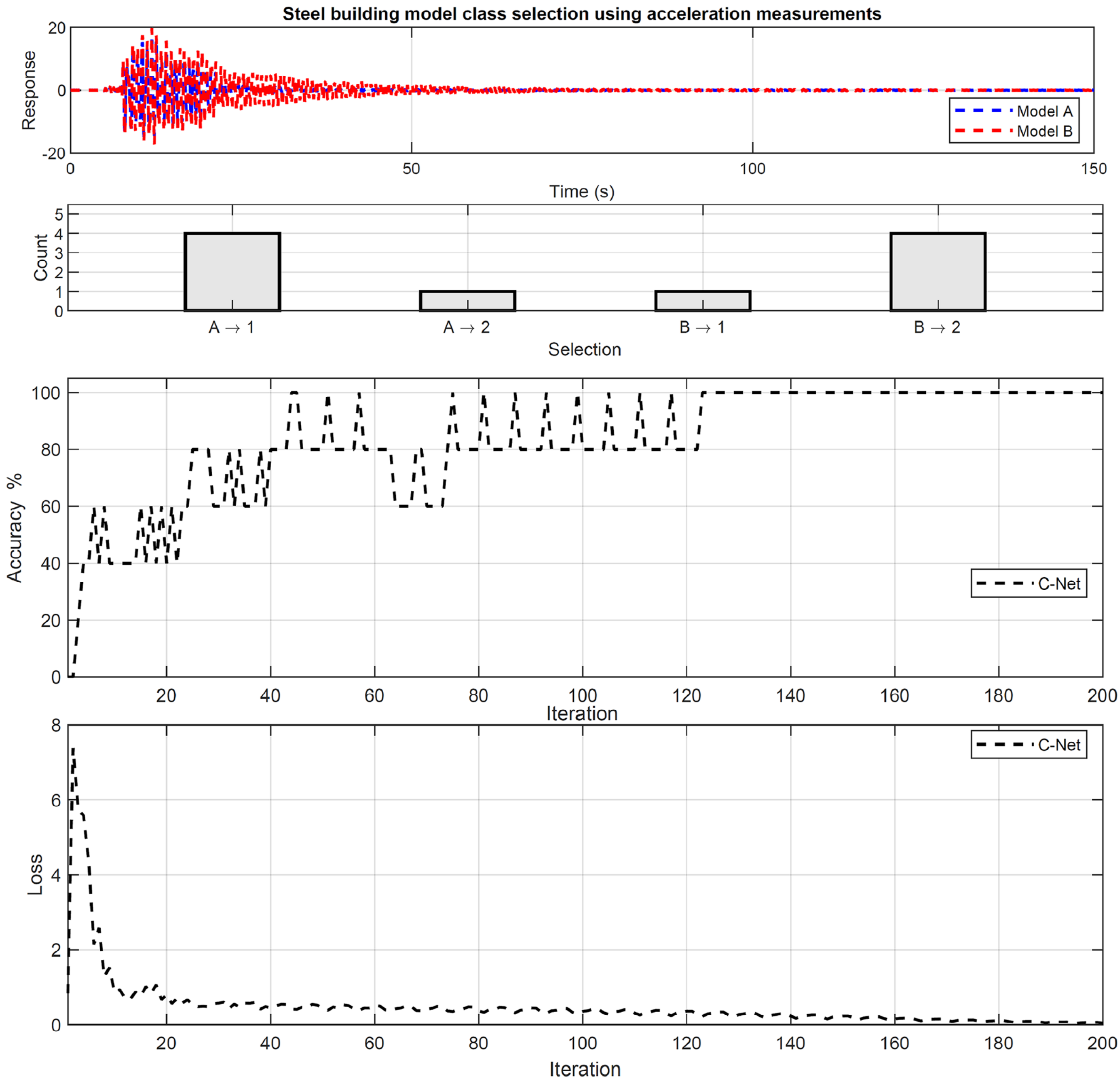

**FIGURE 11** System of Section 8: Results for the 3D building finite element model when training and validating with the top corner DOF acceleration signals (Kobe plot). First row: the displacement raw signals in $m/s^2$. Second row: C-Net model class prediction where ideally A->1 and B->2. Third and fourth row: accuracy and loss in the training process. DOF, degree of freedom.

Relating to the visualization of the results, the horizontal axes of the model class selection may confuse at a first glance. They provide though the prediction of the network relating to the model that the signal was generated, and the model that the signal was classified. In this view, the count number of correct and wrong prediction is seen.

Along these lines, the topic "model class selection" should be clarified better as it touches many engineering fields. In reality, this work did not make any distinction between the field of application, and the potential is open for fields different than the structural identification. For the structural health monitoring field, specifically, the method provides the model that will be further used to identify the structure, without having to perform the identification for each one model first.

Specifically for structural health monitoring applications, the number of candidate models is usually low, and the method manages to provide a reliable prediction. However, for other fields, such as if one wanted to predict a model class for a nonlinear oscillator with some combination of polynomial stiffness terms, one would require $2^n - 1$ candidate model classes to comprehensively consider up to $n^{th}$ order polynomial terms. With regards to this point, future research is recommended for application on those field investigating the number of the models which results in the method to fail, and how the number of candidate models effects the accuracy of the model class predictions. The reason lies into the fact that the number of candidate models would be prone to proliferation in a way that could potentially be detrimental to prediction performance.

Another concern is related to the model class selection capability without the need to identify the parameters. However, for Table 2, it is stated that parameter calibration is performed using system identification techniques, and this seems to be somewhat of a contradiction. In reality, those parameters were used to generate the signals to train the network, and they were not used or identified during the CNN model class selection process.

Regarding the network algorithm parameters, the examinations so far showed a recommendation of as high as possible values for the filter size and the number of neurons in the convolutional layers. The first one defines the kernel where the data are multiplied by, while the second one determines the number of feature maps.

However, this recommendation sounds restrictive or suboptimal since it leads to higher weights for back-propagation, and ultimately to higher computational cost.

Despite this, the computational cost of this approach is bearable. This is attributed to three main reasons: the one-dimensional nature of the data, the unique signal training approach which may be implemented, and the Kalman filtering of the signals to remove the noise.

The higher values for the filter size and the number of neurons recommendation is not mandatory though. The user may achieve the same accuracy with a much lower value of them, and with reduced computational cost. However, for a low number of them, a reduced accuracy is observed despite that the training process wrongly seems to reach a 100% accuracy.

To demonstrate this, consider the examined linear and nonlinear systems. Compared to the previous numerical applications of Sections 6 and 7, only the filter size is changed to 3 and the number of neurons to 8.

Two signal inputs are examined in Figures 12–13 with the same layout description as in Section 6. In total, 9 new velocity and displacement signals are classified for the linear system, and 10 new velocity and displacement signals for the non-linear system. Ideally for the linear dynamic system, the first 3 signals belong to Model A, the second 3 signals belong to Model B, and the last 3 signals belong to Model C. For the free fall nonlinear system, the first 5 signals belong to Model A, and the second 5 signals belong to Model B.

In Figure 12, the performance of the networks in the linear dynamic system using only the DOF 1 displacement signal parts in the training and validation process is shown. Both networks misselect five out of nine signals. Interestingly, both training processes reach a 100% accuracy despite that the loss is high. The loss can be then used as an indication that higher filter size and neural number are needed. Importantly, both networks have nine out nine correct selections for higher filter size and neuron number values, as shown in Section 6.

In Figure 13, the performance of the networks in the nonlinear system using only the DOF 1 displacement signal parts in the training and validation process is shown. Both networks misselect 3 or 4 out of 10 signals. Interestingly, both training processes reach a 100 % accuracy despite that the loss is high. The loss can be then used also in nonlinear systems as an indication that higher filter size and neural number are needed. Importantly, both networks have a higher number of correct class selections for higher filter size and neuron number values, as shown in Section 7.

Here, the sensitivity investigation is performed for a low number of model classes which potentially means that for a larger number of them, larger deviations are expected when the filter size and number of neurons is low. Importantly training the network with multiple number of signals overcomes the inaccuracies derived from low filter size and number of neurons, but increases the computational cost.

Last but not least, the training results and accuracy shows the normal variability of the convolutional neural networks results. In this unique response training approach, this limitation phenomenon is enhanced and additional research is recommended. Importantly, all the applications presented in this work are based on a very limited amount of data for training. In a scenario where a large amount of them (many signals train the network after many earthquake events for the same structure) higher accuracy is expected. However, this is not always available in real-life applications, which led to the low data or unique signal training investigation within this work.

Another concern is related to the investigation into the extrapolation capabilities of the approach since only the outputs measured from a system. The examinations so far showed the potential of the method when the structural model remains the same. However, this assumption may not be true if a change happen to the system, some damage for instance, or any other modification on the structure.

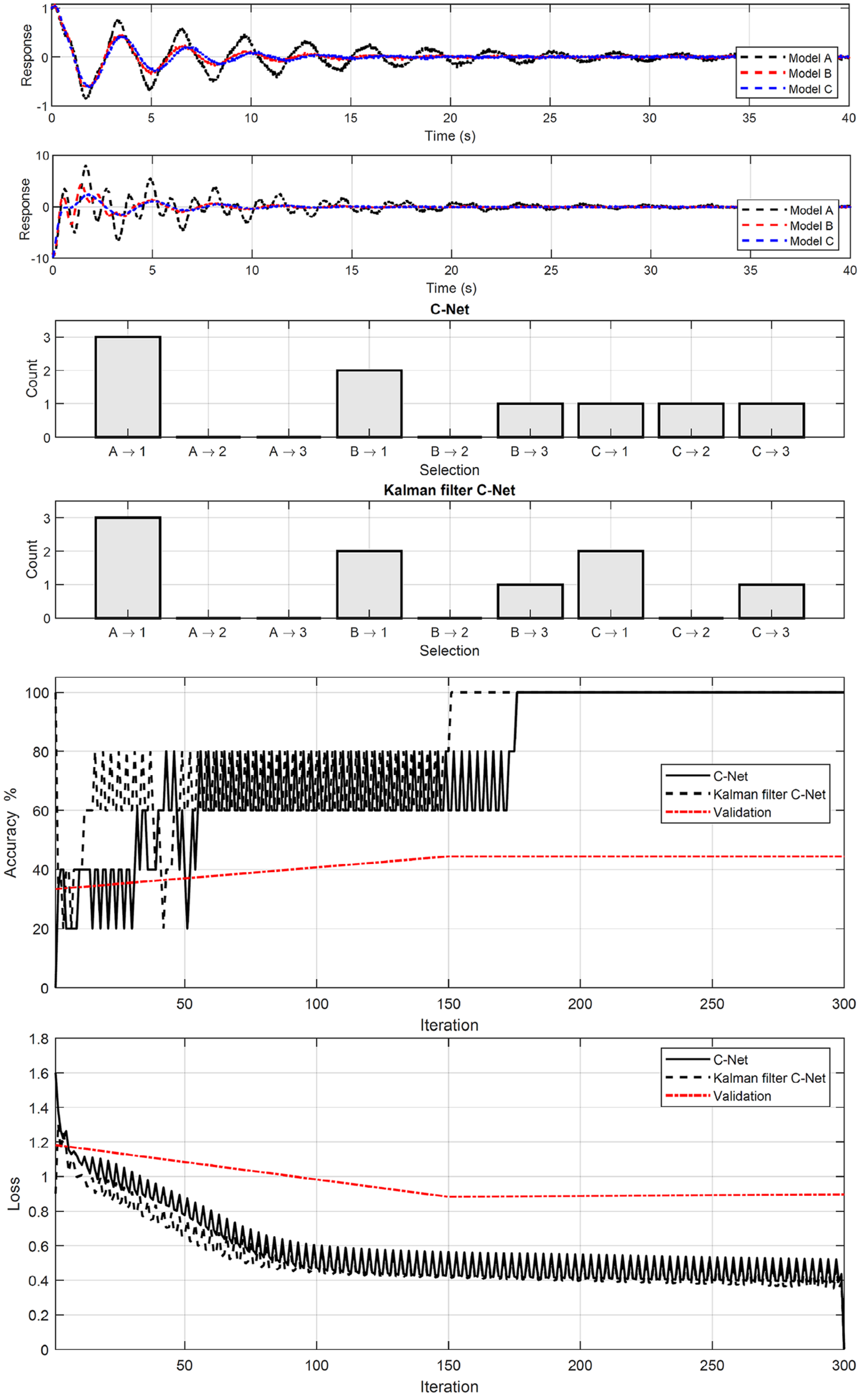

**FIGURE 12** System of Section 6 in discussion Section 9: Results for the linear dynamic system when training and validating with the DOF 1 displacement signals, but with a poor filter size and number of neurons. First and second row: the displacement and acceleration raw signals in m/s and m/s$^2$, respectively. Third row: C-Net class prediction where ideally A->1, B->2, and C->3. Fourth row: Kalman filter C-Net prediction. Fifth and sixth row: accuracy and loss in the training process for both networks. DOF, degree of freedom.

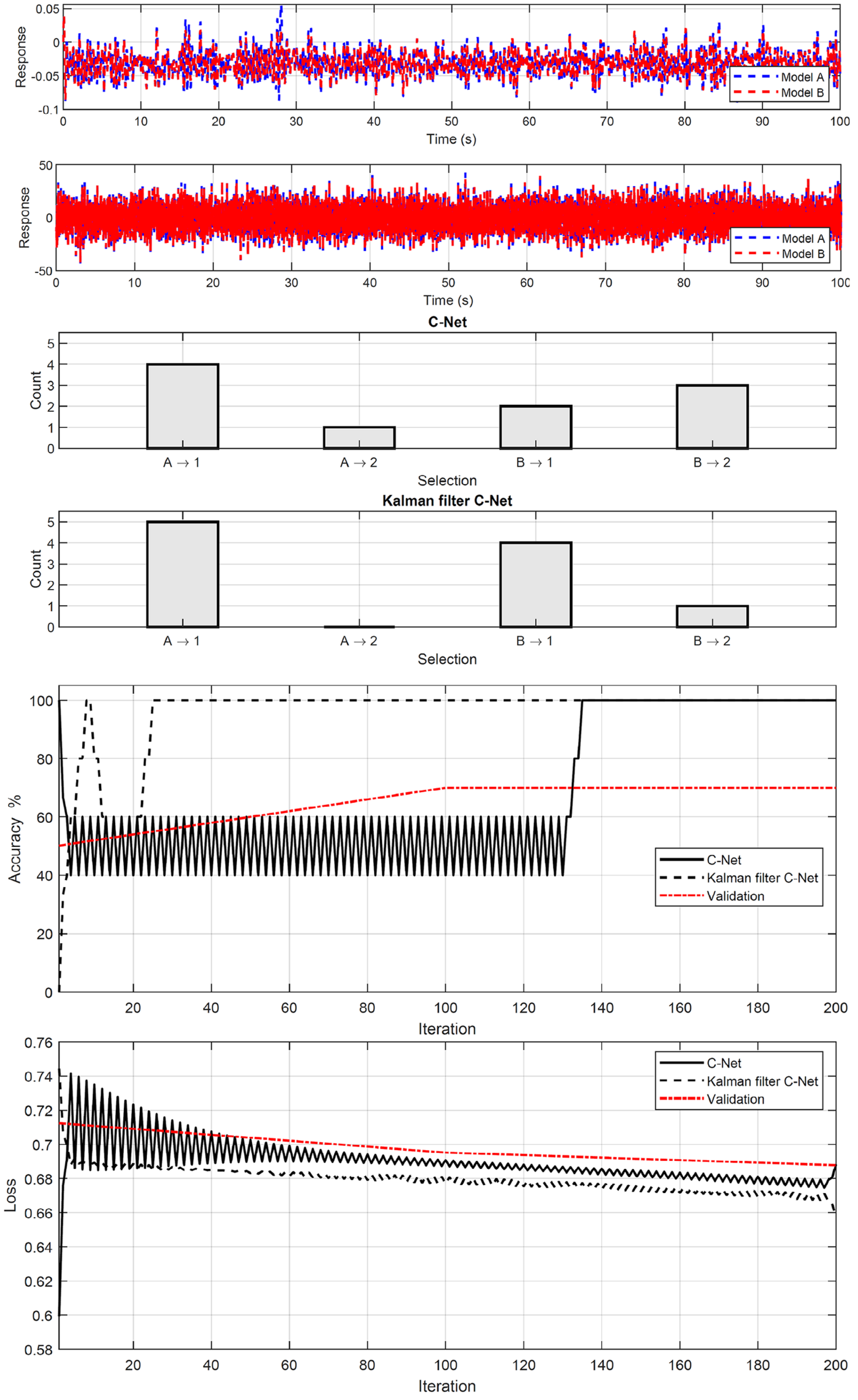

**FIGURE 13** System of Section 7 in discussion Section 9: Results for the nonlinear system when training and validating with the DOF 1 displacement signals, but with a poor filter size and number of neurons. First and second row: the displacement and acceleration raw signals in m/s and m/s$^2$, respectively. Third row: C-Net class prediction where ideally A->1 and B->2. Fourth row: Kalman filter C-Net prediction. Fifth and six row: accuracy and loss in the training process for both networks. DOF, degree of freedom.

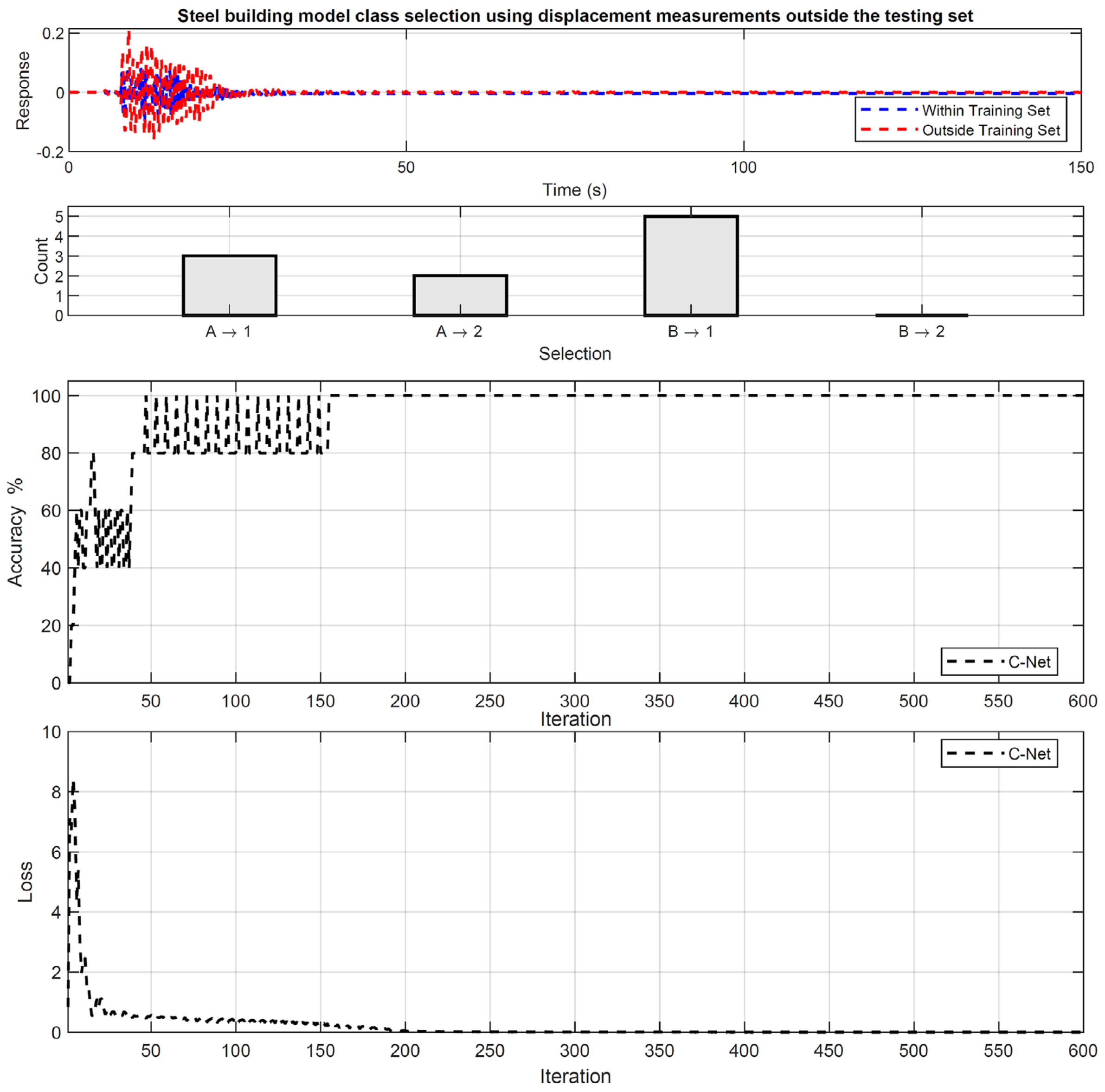

**FIGURE 14** System of Section 8 in discussion Section 9: Results for the 3D building finite element model when training and validating with the top corner DOF displacement signals, but selecting the model class of signals outside the training set (a change on boundary conditions is examined). First row: the displacement raw signals in $m$. Second row: C-Net model class prediction where ideally A->1 and B->2. Third and fourth row: accuracy and loss in the training process. DOF, degree of freedom.

To explore this, consider the examined 3D building. Compared to the previous numerical applications of Section 8, only some of the ground boundary conditions are changed to allow rotation, instead of fixed nodes (termed “outside training set” response in Figures 14–16). This simulates a damage scenario at the foundation of the structure, for instance.

Three signal inputs are examined in Figures 14–16 with the same layout description as in Section 8. In total, 10 new displacement, velocity, and acceleration signals are classified, where ideally the first 5 signals belong to Model A, and the second 5 signals belong to Model B.

In Figure 14, the performance of the network is shown using only the top corner building DOF displacement signals. The C-Net misselects 7 out of 10 signals. Compared to Figures 12–13, both training processes reach a 100 % accuracy and the loss is low. The loss, then, cannot be used as an indication that the prediction is wrong. The same conclusion is derived in Figure 15 and 16 for the performance of the network using only the top corner building DOF velocity or acceleration signals, respectively.

As a result, the approach is not capable of some form of extrapolation to predict model classes for systems with forcings outside of the training dataset to ensure good performance. When employed on a real engineering system where the system may change, one must have some prior belief about the expected forcing patterns in order to generate comprehensive training datasets, and retrain the network for future good prediction. It follows, as a future recommendation, that one requires some prior belief regarding anticipated forcings in order to use the approach, and the method can be combined with respect to Bayesian model selection approaches with Bayesian latent force estimation.[96,97] This is a pertinent test for model class selection approaches in engineering applications as there could be high-cost or safety critical ramifications if a model class is confidently predicted incorrectly.

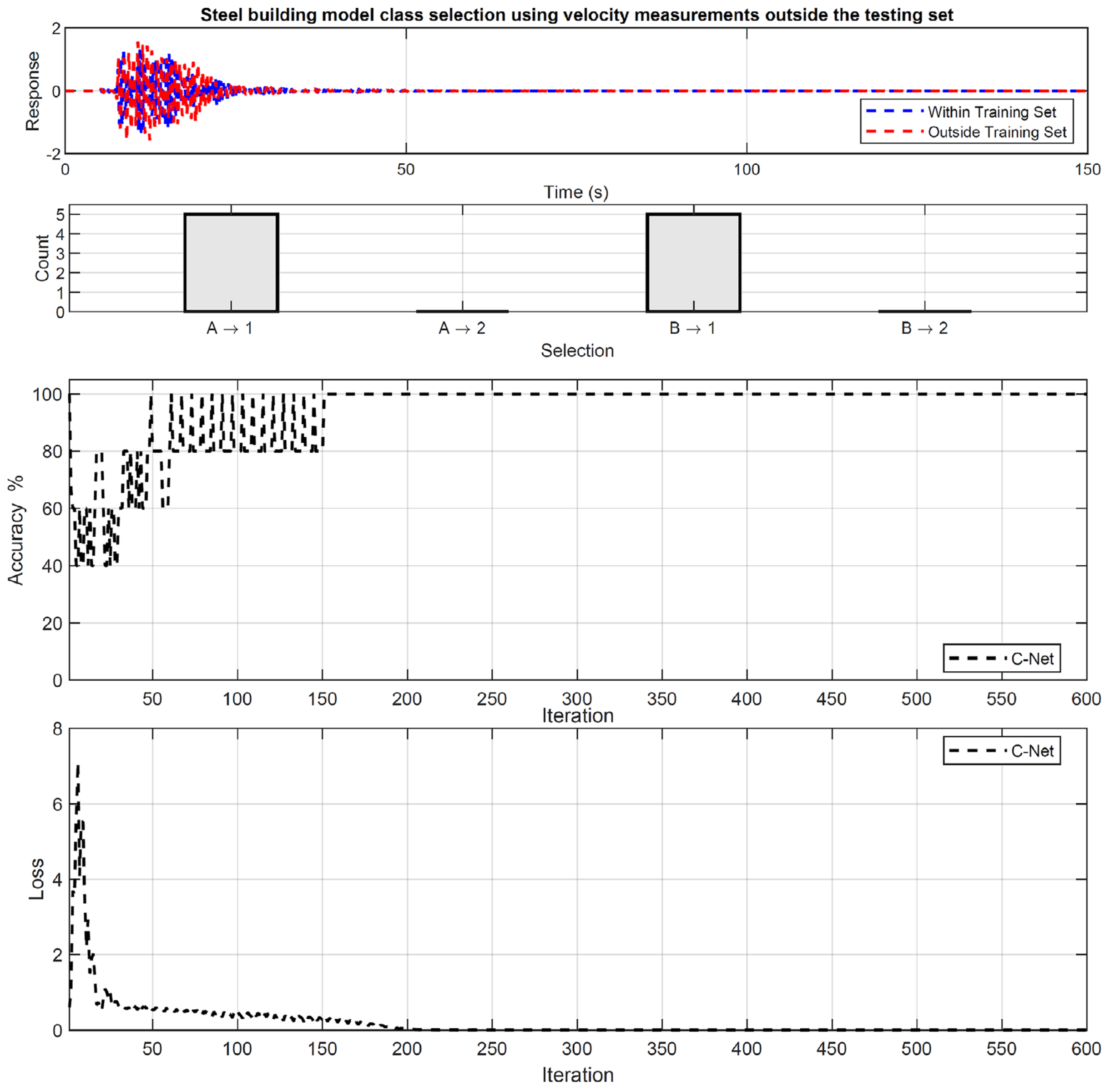

**FIGURE 15** System of Section 8 in discussion Section 9: Results for the 3D building finite element model when training and validating with the top corner DOF velocity signals, but selecting the model class of signals outside the training set (a change on boundary conditions is examined). First row: the displacement raw signals in m/s. Second row: C-Net model class prediction where ideally A->1 and B->2. Third and fourth row: accuracy and loss in the training process. DOF, degree of freedom.

A final concern is related to the uncertainty quantification where the model class selection methodology should provide. Namely, a desirable property for model class prediction approaches to possess that accurately representing the uncertainty around predictions. In the framework of convolutional neural networks, this may achieved by retraining the model multiply times and take the average and the rest statistical properties of the network prediction.

Last but not least, regarding using other types of neural networks such as the long short-term memory ones,[98] an investigation was made. The long short-term memory neural networks are widely recognized as a powerful machine learning tool for both classification and regression problems. They belong to the wider library of the recurrent neural networks which use feedback loops with recurrent connections between the nodes of the network to make them capable of modeling sequences of signals, such as the structural vibration raw signal $\mathbf{y}$.

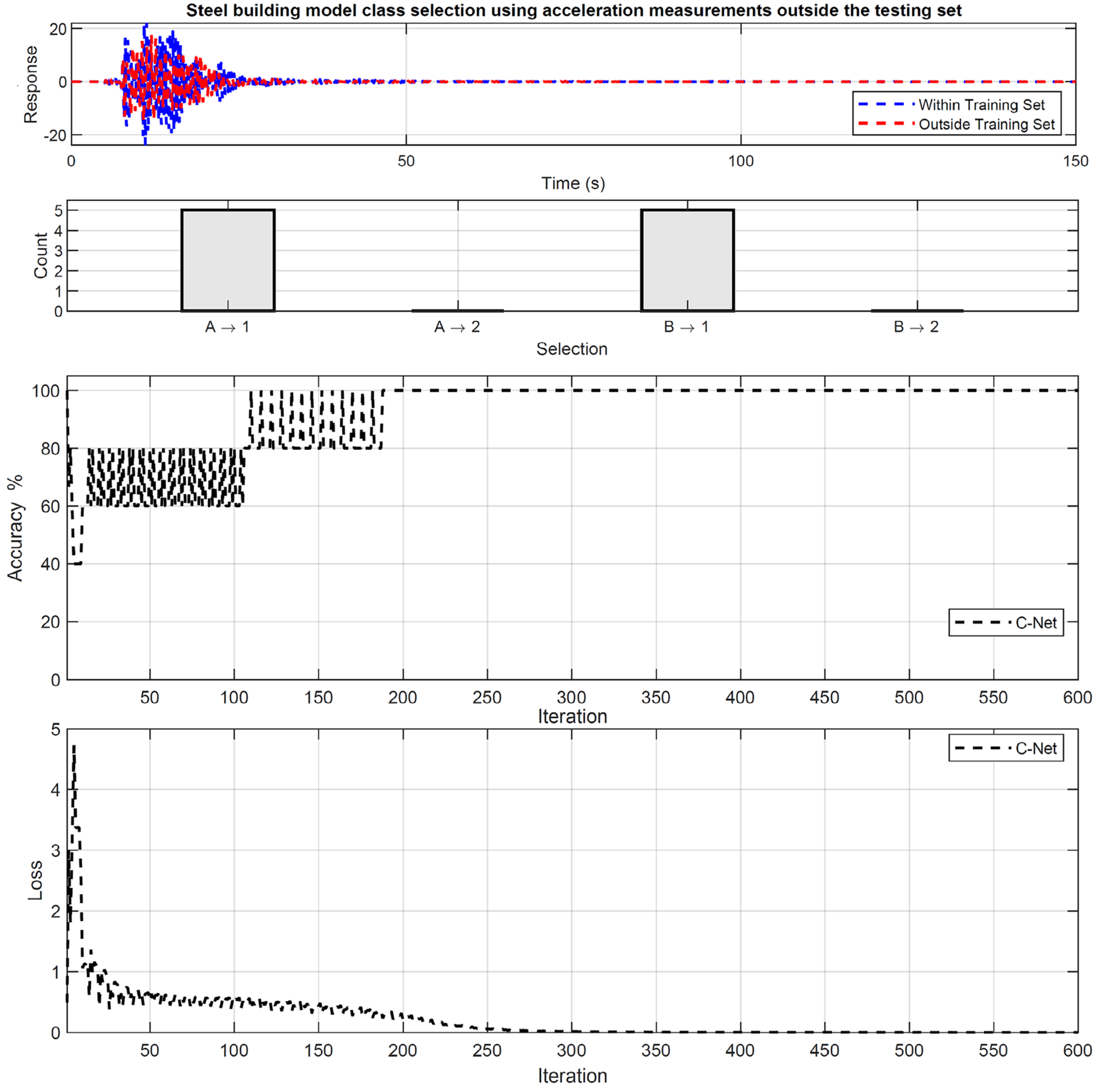

**FIGURE 16** System of Section 8 in discussion Section 9: Results for the 3D building finite element model when training and validating with the top corner DOF acceleration signals, but selecting the model class of signals outside the training set (a change on boundary conditions is examined). First row: the displacement raw signals in m/s$^2$. Second row: C-Net model class prediction where ideally A->1 and B->2. Third and fourth row: accuracy and loss in the training process. DOF, degree of freedom.

The intuition behind the them is to create an additional module in a neural network that learns when to remember and when to forget some characteristic of the provided vibration signal. In other words, the network, effectively learns which patterns might be needed in the signal and when that information is no longer needed. This poses an advantage for structural model selection among a group $\mathbb{M}$ of them when an unexpected excitation excites the structure which, as not attributed to model response to ambient environment, does not play an important role in the final model selection, and can be neglected. Importantly, this unexpected excitation is potentially of unknown magnitude, and the network does not need to have this information to perform the model selection.

The discussed, though, long short-term memory gates make the training more difficult and increase the training time of the network. To reduce training time and improve network performance, a simplified but improved gated recurrent unit architecture network[99] may also introduced for structural model selection. The gated recurrent unit chooses a new type of hidden unit that merges the forget gate and the vibration signal **y** gate into a single update gate, and mixes also the cellular and the hidden state into one state. The number of gates is decreased compared to long short-term memory which are termed update gate and reset gates. The final model is simpler than the standard long short-term memory resulting in a faster convergence for structural health monitoring applications. The author switched both convolutional layers with long short-term memory and gated recurrent unit ones, keeping the same architecture, and both of them always underperformed the convolutional one architecture. Additional research is therefore recommended on how those layers and architectures can compete the convolutional one in model class selection problems.

Finally, future directions are also provided in the area of using clustering techniques to judge which model class a signal belongs to, if it provides in an easier way to solve this problem, and what are the limitations compared to this work. Importantly though, the clustering approach does not incorporate a labeling philosophy to associate the signals to some models.

## 10 | CONCLUSIONS

The response-only model class selection capability of a novel deep convolutional neural network method was examined herein in a simple, yet effective, manner. Specifically, the responses from a unique DOF along with their class information trained and validated a one-dimensional convolutional neural network. In doing so, the network selected the model class of new and unlabeled signals without the need of the system input information, or full system identification. An optional physics-based algorithm enhancement was also examined using the Kalman filter to fuse the system response signals using the kinematics constraints of the acceleration and displacement data.

Overall, this method allowed for the model class selection with:

1. Real-time application when the network has been trained.
2. Automatic and response-only outcome without the need of the system input information.
3. A unique DOF application without full system identification, or the dynamic state estimation of potentially partially unobservable systems.
4. The absent of a strict mathematical representation of the system nonlinear behavior.
5. The use of filtered signals instead of the common approach with raw-data in convolutional neural networks.
6. Independent to the system type application.

Importantly, the method was shown to select the model class in slight signal variations attributed to the damping behavior or hysteresis behavior on both linear and nonlinear dynamic systems, as well as on a 3D building finite element model, providing a powerful tool for structural health monitoring applications. Related to the limitations, this approach does not provide information on the system input, parameter and dynamic state estimation, while it is also vulnerable to the proper training in a region close the unknown model.

## ACKNOWLEDGMENTS

The author would like to gratefully acknowledge the reviewers for their constructive comments, John T. Katsikadelis for the discussion on benchmark integro-differential equation problems, and Andrew W. Smyth for the previous insightful discussions on model class selection and Kalman filtering.

## DATA AVAILABILITY STATEMENT

The data that support the findings of this study are available from the corresponding author upon reasonable request.